\documentclass{SciPost}
\usepackage[T1]{fontenc}
\usepackage[bitstream-charter]{mathdesign}
\usepackage{booktabs,array}
\usepackage{float,needspace}
\usepackage[column=G]{cellspace}
\DeclareSymbolFont{usualmathcal}{OMS}{cmsy}{m}{n}
\DeclareSymbolFontAlphabet{\mathcal}{usualmathcal}
\hypersetup{colorlinks,linkcolor={red!50!black},citecolor={blue!50!black},urlcolor={blue!80!black}}
\fancypagestyle{SPstyle}{
  \fancyhf{}
  \lhead{\colorbox{scipostblue}{\bfseries\color{white}~SciPost Physics~}}
  \rhead{\bfseries\color{scipostdeepblue}Submission}
  
  \fancyfoot[C]{\textbf{\thepage}}
}

\newtheorem{theorem}{Theorem}
\newtheorem{lemma}{Lemma}
\newtheorem{corollary}{Corollary}
\newtheorem{proposition}{Proposition}
\newtheorem{appendixlemma}{Lemma}[section]
\hypersetup{pdftitle={Hybrid correlation generation with finite shared randomness},
  pdfauthor={Zihan Chen, Shengjun Wu, and Zeng-Bing Chen}}

\begin{document}

\pagestyle{SPstyle}

\begin{center}{\Large\bfseries\color{scipostdeepblue} Hybrid correlation generation with finite shared randomness}\end{center}

\begin{center}\textbf{Zihan Chen\textsuperscript{1}, Shengjun Wu\textsuperscript{1$\star$} and Zeng-Bing Chen\textsuperscript{1$\dagger$}}\end{center}
\begin{center}\textbf{1} National Laboratory of Solid State Microstructures and School of Physics, Collaborative Innovation Center of Advanced Microstructures, Nanjing University, Nanjing 210093, China\\
$\star$ \href{mailto:sjwu@nju.edu.cn}{\small\nolinkurl{sjwu@nju.edu.cn}}\,,\quad
$\dagger$ \href{mailto:zbchen@nju.edu.cn}{\small\nolinkurl{zbchen@nju.edu.cn}}\end{center}

\section*{\color{scipostdeepblue}Abstract}
{\bfseries\boldmath Finite shared randomness can reduce the use of a prescribed two-qubit source in a hybrid model. For
two binary projective measurements per party, exact simulation uses private randomness to choose
classical or quantum interfaces. The sharp bounds are four labels for minimum source cost and eight
for all feasible pairs of interface choice probabilities. For Werner states with identical \(X/Z\)
measurements, exact formulas cover all label budgets. The three-label cost jumps from zero as
visibility exceeds \(\frac{2}{3}\). In a separable family, the two-label cost is nonmonotonic in
local bias at fixed correlation tensor. The minimum worst-case total variation error is derived for
two-label classical sampling.}

\vspace{\baselineskip}
\noindent\textcolor{white!90!black}{%
\fbox{\parbox{\dimexpr\linewidth-2\fboxsep-2\fboxrule\relax}{%
\textcolor{white!40!black}{\begin{tabular}{@{}l@{}r@{}}%
  \begin{minipage}{0.59\linewidth}%
    {\small Copyright attribution to authors. \newline
    This work is a submission to SciPost Physics. \newline
    License information to appear upon publication. \newline
    Publication information to appear upon publication.}
  \end{minipage} & \begin{minipage}{0.39\linewidth}
    {\small Received Date \newline Accepted Date \newline Published Date}%
  \end{minipage}
\end{tabular}}
}}
}

\nolinenumbers
\setcounter{tocdepth}{1}
\vspace{10pt}\noindent\rule{\textwidth}{1pt}
\tableofcontents
\noindent\rule{\textwidth}{1pt}\vspace{10pt}

\flushbottom

\titlespacing*{\subsection}{0pt}{1.0\baselineskip}{0.4\baselineskip}
\titlespacing*{\subsubsection}{0pt}{0.85\baselineskip}{0.35\baselineskip}

\section{Introduction}
\label{sec:1}

The exact generation of prescribed correlations between spatially separated parties is a basic task in quantum
information \cite{Brunner2014}. Bell-local correlations admit local hidden-variable (LHV) models with
input-independent shared randomness and no communication \cite{Fine1982}. Such models also reproduce joint
statistics of local measurements on certain entangled states \cite{Werner1989}. While resource theories based
on local operations and shared randomness (LOSR) treat shared randomness as free
\cite{deVicente2014,Schmid2020,Rosset2020}, bounding the support size of the shared variable imposes a
shared-label budget. Existing results provide upper bounds on the minimum number of labels required for exact
simulation \cite{Rosset2018,Bowles2015}. Restricting this number can make the attainable correlation sets
nonconvex and reveal dimensional advantages of quantum realizations \cite{Donohue2015,deVicente2017}. For
example, certain Bell-local correlations generated by two-qubit systems require more than two shared labels in
a local classical model, a phenomenon known as superlocality \cite{Jebaratnam2017}.

Shared randomness requirements also arise in local hidden-state (LHS) models
\cite{Das2018,Jebarathinam2019,Zhang2025}. Each label specifies a classical response at one party and a
quantum state at the other. For some unsteerable correlations, LHV--LHS simulation requires more shared labels
than the untrusted party's local Hilbert-space dimension in a quantum realization. This is termed
superunsteerability \cite{Das2018}. For a given assemblage, the steering-equivalent measurements are defined
on the support of the trusted party's reduced state. If the assemblage is unsteerable, these measurements admit a common parent
measurement \cite{Quintino2014,Uola2015}. With stochastic postprocessing, the minimum number of parent
outcomes equals the minimum number of LHS labels required to reproduce that assemblage
\cite{Zhang2025,Skrzypczyk2020}.

Classical--quantum resource tradeoffs have also been studied in fixed-dimensional correlation generation
\cite{Guha2021} and two-stage hybrid protocols \cite{Lin2022}. Complementary quantities, including local
content, steerable weight, and steering cost, quantify component weights in convex decompositions of
correlations or assemblages \cite{Barrett2006,Skrzypczyk2014,DasSteering2018}. A local behavior may require
more shared labels for exact simulation than the budget allows. This raises a resource question: What is the
minimum source cost when classical and one-sided quantum implementations are also available? The task
considered here fixes a two-qubit state, two binary projective measurements per party, and a common
shared-label budget.

\begin{figure}[tbp]
  \centering
  \includegraphics[width=\linewidth]{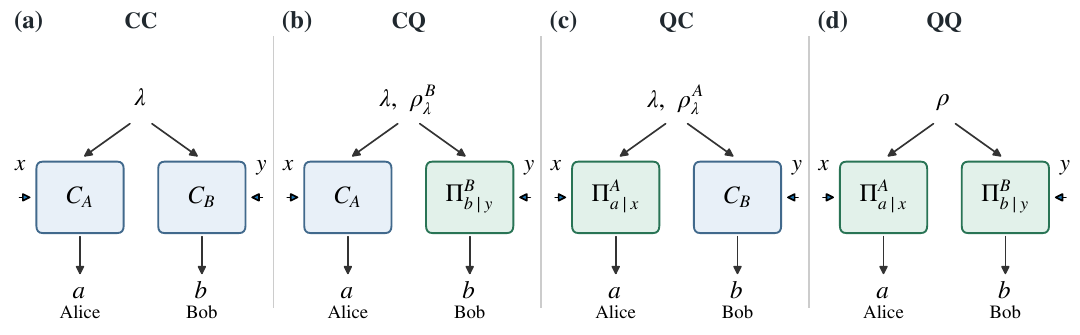}
  \caption{Hybrid correlation generation with classical (C) and quantum (Q) interfaces. The parties choose their
interfaces independently of each other and of the shared label before receiving inputs. The CC, CQ, and QC
branches share the same label distribution and reuse the same local classical response functions. Quantum
interfaces apply the fixed local projective measurements to label-dependent qubit states in CQ and QC, or to
the prescribed bipartite state \textit{\(\rho\)} in QQ.}
  \label{fig:1}
\end{figure}

Before receiving inputs, the parties independently select classical (C) or quantum (Q) interfaces and inform
the preparation device (see Fig.~\ref{fig:1}). Their choices are independent of the shared label. For example,
each controller can use private randomness to select its interface before the preparation device draws the
shared label from a fixed distribution. The device supplies label-conditioned local qubit states in the CQ and
QC branches, while the QQ branch uses the prescribed bipartite state. Each party's classical response depends
only on its local input and label, with the same response function reused across branches. Each quantum
interface performs the prescribed measurement for its input. The CC, CQ, and QC branches share one label
distribution. The QQ output is independent of the label. No communication with the other party or the device
is allowed after input arrival. The source cost of each protocol is its QQ-branch probability. The task is to
minimize this probability over exact protocols within the shared-label budget.

Eight shared labels suffice to realize every feasible pair of interface choice probabilities, and this bound
is tight. The classical four-label bound \cite{Donohue2015} implies that four shared labels attain the
unrestricted minimum source cost, which is zero for local behaviors and one otherwise. For Werner states with
the same Pauli \(X\) and \(Z\) measurements on both sides, exact formulas give the minimum source
cost for every shared-label budget. With three shared labels, the cost remains zero up to and including a
visibility of \(\frac{2}{3}\). Its right-hand limit there is approximately 0.805. With stochastic postprocessing
of a parent measurement having at most three outcomes, the maximum visibility for noisy Pauli \(X\) and \(Z\)
measurements is approximately 0.5609. The construction in Ref.~\cite{Skrzypczyk2020} attains this value.
A family of separable states exhibits a nonmonotonic two-label cost as local bias varies at fixed
correlation tensor. Flipping Bob's output transfers the Werner costs to the sampling task in Sec.~\ref{sec:7}. For classical sampling
with at most two shared labels, matching upper and lower bounds determine the minimum worst-case total
variation error as a function of the source flip probability.

\section{Model and resource constraints}
\label{sec:2}

Consider two spatially separated parties, Alice and Bob. After resource preparation, Alice and Bob receive inputs
\(x,y\in \{0,1\}\) and produce outputs \(a,b\in \{-1,+1\}\) without further communication. The task is to
reproduce the target conditional distribution \(P(a,b|x,y)\) under the prescribed resource constraints.

\Needspace{3.5\baselineskip}
Fix a two-qubit state \(\rho\) and local measurements:
\begin{equation*}
\mathcal{M}=(A_{0},A_{1},B_{0},B_{1}),
\end{equation*}
where
\begin{equation*}
A_{x}=\mathbf{a}_{x} \cdot \boldsymbol{\sigma},\quad B_{y}=\mathbf{b}_{y} \cdot \boldsymbol{\sigma}.
\end{equation*}
Here \(\boldsymbol{\sigma}=(\sigma_{x},\sigma_{y},\sigma_{z})\) is the vector of Pauli matrices, with
\(X=\sigma_{x}, Y=\sigma_{y}, Z=\sigma_{z}.\) The unit vectors \(\mathbf{a}_{x},\mathbf{b}_{y} \in
\mathbb{R}^{3}\) specify the measurement directions. The corresponding projectors are
\begin{equation*}
\Pi_{a|x}^{A}=\frac{I+aA_{x}}{2},\quad \Pi_{b|y}^{B}=\frac{I+bB_{y}}{2}.
\end{equation*}
\(I_{n}\) denotes the \textit{n}-dimensional identity and \(I=I_{2}\). The target distribution is
\begin{equation}
\label{eq:1}
P(a,b|x,y)=\operatorname{Tr} \,\left[ \rho \left( \Pi_{a|x}^{A} \otimes \Pi_{b|y}^{B}\right)\right].
\end{equation}
Define the local means and correlators by
\begin{subequations}
\begin{equation}
\label{eq:2a}
m_{Ax}=\operatorname{Tr} \,[\rho (A_{x} \otimes I)],\quad m_{By}=\operatorname{Tr} \,[\rho (I\otimes B_{y})],
\end{equation}

\begin{equation}
\label{eq:2b}
E_{xy}=\operatorname{Tr} \,[\rho (A_{x} \otimes B_{y})].
\end{equation}
\end{subequations}
For binary outcomes, the local means determine the local marginals. The local mean vectors and correlation
matrix are
\begin{equation*}
m_{A}=(m_{A0},m_{A1})^{\mathsf{T}},\;m_{B}=(m_{B0},m_{B1})^{\mathsf{T}},\quad E=(E_{xy})_{x,y=0,1},
\end{equation*}
where the superscript \(\mathsf{T}\) denotes transposition. These eight moments completely determine the target behavior:
\begin{equation}
\label{eq:3}
P(a,b|x,y)=\frac{1}{4}\left( 1+am_{Ax}+bm_{By}+abE_{xy}\right).
\end{equation}
In a \(K\)-label protocol, the shared classical variable \(\Lambda\) takes at most \(K\) values with
positive probability. The label probabilities satisfy
\begin{equation*}
p_{\lambda}=\Pr (\Lambda=\lambda),\qquad  p_{\lambda} \ge 0,\quad \sum_{\lambda=1}^{K} p_{\lambda}=1.
\end{equation*}
Conditioned on \(\Lambda\), the parties may use independent private randomness. All shared randomness used by
the protocol is included in \(\Lambda\).

Before receiving their inputs, Alice and Bob independently choose a classical (C) or quantum (Q) interface.
Their joint choice is independent of \(\Lambda\). The preparation device supplies the corresponding resources.
Let \(\eta_{A}, \eta_{B} \in [0,1]\) denote their respective probabilities of choosing C. The probability pair
\((\eta_{A}, \eta_{B})\) is called a mode pair. The four branch probabilities are listed in Table~\ref{tab:1}.
These weights satisfy
\begin{equation}
\label{eq:4}
w+\beta+\gamma+\delta=1,\quad w\delta=\beta \gamma.
\end{equation}

\begingroup
\setlength{\intextsep}{6pt}
\begin{table}[H]
  \centering
  \captionsetup{skip=6pt}
  \caption{Branches and probabilities for independent local interface choices.}
  \label{tab:1}
  \small
  \setlength{\tabcolsep}{6pt}
  \renewcommand{\arraystretch}{1.15}
  \begin{tabular}{|Gc|Gc|Gc|Gc|}
    \hline
    \textbf{Branch} & \textbf{Alice interface} & \textbf{Bob interface} & \textbf{Branch probability} \\
    \hline
    CC & C & C & \(w=\eta_{A} \eta_{B}\) \\
    \hline
    CQ & C & Q & \(\beta=\eta_{A} (1-\eta_{B})\) \\
    \hline
    QC & Q & C & \(\gamma=(1-\eta_{A})\eta_{B}\) \\
    \hline
    QQ & Q & Q & \(\delta=(1-\eta_{A})(1-\eta_{B})\) \\
    \hline
  \end{tabular}
\end{table}
\endgroup
The input pair \((x,y)\) is drawn independently of the parties' interface choices, the shared label \(\Lambda\), and the prepared states.
For fixed inputs, the QQ outcome distribution is independent of the shared label. The CC, CQ, and QC branches share the label distribution \(p_\lambda\). At a classical interface, the local input and the realized label value \(\lambda\) determine the outcome probabilities.
\begin{equation}
\label{eq:5}
C_{A} (a|x,\lambda)=\frac{1+au_{\lambda x}}{2},\quad C_{B} (b|y,\lambda)=\frac{1+bv_{\lambda y}}{2}.
\end{equation}
Here \(u_{\lambda x},v_{\lambda y} \in [-1,1]\) are the conditional response means. Define
\(u_{\lambda}=(u_{\lambda 0},u_{\lambda 1})^{\mathsf{T}}\) and \(v_{\lambda}=(v_{\lambda 0},v_{\lambda
1})^{\mathsf{T}}\). Alice uses the same \(u_{\lambda}\) in CC and CQ, while Bob uses the same \(v_{\lambda}\)
in CC and QC.

The QC and CQ branches use label-dependent qubit states \(\rho_{\lambda}^{A}\) and \(\rho_{\lambda}^{B},\)
respectively. Both pure and mixed states are allowed. The prescribed measurements give
\begin{subequations}
\begin{equation}
\label{eq:6a}
Q_{A} (a|x,\lambda)=\operatorname{Tr} \,[\rho_{\lambda}^{A} \Pi_{a|x}^{A}]=\frac{1+ar_{\lambda x}}{2},
\end{equation}

\begin{equation}
\label{eq:6b}
Q_{B} (b|y,\lambda)=\operatorname{Tr} \,[\rho_{\lambda}^{B} \Pi_{b|y}^{B}]=\frac{1+bs_{\lambda y}}{2},
\end{equation}
\end{subequations}
where
\begin{equation*}
r_{\lambda x}=\operatorname{Tr} \,[\rho_{\lambda}^{A} A_{x}],\quad s_{\lambda y}=\operatorname{Tr} \,[\rho_{\lambda}^{B} B_{y}].
\end{equation*}
All response vectors are column vectors. Unsubscripted vector norms are Euclidean. For each party, the allowed
quantum response vectors form a linear image of the Bloch ball. Orthogonal measurement pairs give the unit disks
\begin{equation}
\label{eq:7}
\Vert r_{\lambda} \Vert_{2} \le 1,\quad \Vert s_{\lambda} \Vert_{2} \le 1.
\end{equation}
The Werner-state analysis assumes orthogonal pairs at each party. Alice's and Bob's pairs need not coincide.

The non-QQ branch behaviors are
\begin{subequations}
\begin{equation}
\label{eq:8a}
P_{\mathrm{CC}} (a,b|x,y)=\sum_{\lambda} p_{\lambda} C_{A} (a|x,\lambda)C_{B} (b|y,\lambda),
\end{equation}

\begin{equation}
\label{eq:8b}
P_{\mathrm{CQ}} (a,b|x,y)=\sum_{\lambda} p_{\lambda} C_{A} (a|x,\lambda)Q_{B} (b|y,\lambda),
\end{equation}

\begin{equation}
\label{eq:8c}
P_{\mathrm{QC}} (a,b|x,y)=\sum_{\lambda} p_{\lambda} Q_{A} (a|x,\lambda)C_{B} (b|y,\lambda).
\end{equation}
\end{subequations}
By Eq.~\eqref{eq:1}, the QQ branch reproduces the target distribution.
\begin{equation}
\label{eq:9}
P_{\mathrm{QQ}}(a,b|x,y)=P(a,b|x,y).
\end{equation}
Exact simulation requires
\begin{equation}
\label{eq:10}
P_{\mathrm{hyb}}=wP_{\mathrm{CC}}+\beta P_{\mathrm{CQ}}+\gamma P_{\mathrm{QC}}+\delta P=P.
\end{equation}
Both the overbar and \(\mathbb{E}\) denote averaging over the shared label. For example,
\begin{equation*}
\overline{u}=\mathbb{E}[u_{\Lambda}]=\sum_{\lambda} p_{\lambda} u_{\lambda},\overline{uv^{\mathsf{T}}}=\sum_{\lambda} p_{\lambda} u_{\lambda} v_{\lambda}^{\mathsf{T}}.
\end{equation*}
Let \(\tau=1-\delta.\) Exact simulation is equivalent to
\begin{subequations}
\begin{align}
\label{eq:11a}
&\eta_{A} \overline{u}+\gamma \overline{r}=\tau m_{A},\\
\label{eq:11b}
&\eta_{B} \overline{v}+\beta \overline{s}=\tau m_{B},\\
\label{eq:11c}
&w\overline{uv^{\mathsf{T}}}+\beta \overline{us^{\mathsf{T}}}+\gamma \overline{rv^{\mathsf{T}}}=\tau E.
\end{align}
\end{subequations}
For fixed \(\left( \rho,\mathcal{M}\right)\), let \(\mathcal{F}_{K} (P)\) be the set of mode pairs admitting
an exact protocol with at most \(K\) shared labels. The source cost of a protocol is its QQ-branch
probability \(\delta\). The minimum source cost at budget \(K\) is
\begin{equation}
\label{eq:12}
q_{K} (P)=\min_{(\eta_{A},\eta_{B})\in \mathcal{F}_{K} (P)} \,(1-\eta_{A})(1-\eta_{B}).
\end{equation}
The mean number of quantum-interface activations per round is \(N_{\mathrm{Q}}=(1-\eta_{A})+(1-\eta_{B}).\)

Relaxing the requirement that the parties choose their interfaces independently can reduce the source cost.
For identical Pauli \(X\) and \(Z\) measurements, let a uniform shared sign \(\lambda=\pm 1\) supply the two
labels. Suppose an additional shared bit, independent of \(\lambda\) and supplied at no cost, selects CQ or QC
with equal probability. In CQ take \(u_{\lambda}=-s_{\lambda}=\left( \lambda,0\right)^{\mathsf{T}}\), and in
QC take \(r_{\lambda}=-v_{\lambda}=\left( 0,\lambda\right)^{\mathsf{T}}\). The local means vanish, and
\(E=-\frac{I_{2}}{2}\) with \(\delta=0\). With independent interface choices, Eq.~\eqref{eq:24} gives \(q_2\!\left(\tfrac{1}{2}\right)>0\) for the same target.

\section{General label-capacity structure and saturation}
\label{sec:3}

Since zero-weight labels are allowed, every protocol feasible with at most \(K\) shared labels remains
feasible with \(K+1\) labels. Hence,
\begin{equation}
\label{eq:13}
\mathcal{F}_{K} \left( P\right) \subseteq \mathcal{F}_{K+1} \left( P\right),\quad q_{K+1} \left( P\right) \le q_{K} \left( P\right).
\end{equation}
Define the unrestricted feasible mode set and optimal value by
\begin{equation}
\label{eq:14}
\mathcal{F}_{\infty} \left( P\right)=\bigcup_{K\ge 1} \mathcal{F}_{K} \left( P\right),\quad q_{\infty} \left( P\right)=\inf_{K\ge 1} \,q_{K} \left( P\right).
\end{equation}
Cost saturation means \(q_{K} (P)=q_{\infty} (P),\) while feasible-set saturation means \(\mathcal{F}_{K}
(P)=\mathcal{F}_{\infty} (P)\). The latter may require more shared labels.

\subsection{Locality implied by the fixed QQ branch}
\label{sec:3.1}

Let \(\mathcal{L}\) denote the set of local behaviors with no bound on the number of shared labels.

\begin{lemma}
\label{lem:1}
\normalfont
If a feasible protocol has \(\delta<1\), then \(P\in \mathcal{L}\).
\end{lemma}

\textit{Proof.} For \(\delta<1\), Eq.~\eqref{eq:10} expresses \(P\) as a mixture of the CC, CQ, and QC behaviors,
\begin{equation*}
P=\frac{wP_{\mathrm{CC}}+\beta P_{\mathrm{CQ}}+\gamma P_{\mathrm{QC}}}{1-\delta}.
\end{equation*}
Each branch behavior is local, so their normalized mixture is local. This decomposition may use more than
\(K\) labels because the shared variable also records the branch type.

\subsection{Zero QQ probability and pure CC realizations}
\label{sec:3.2}

\begin{lemma}
\label{lem:2}
\normalfont
Within the model of Sec.~\ref{sec:2}, assume no further restrictions on \(\eta_{A},\eta_{B}\). Then \(q_{K}
\left( P\right)=0\) if and only if \(P\) admits a pure-CC realization with at most \(K\) shared labels.
\end{lemma}

\textit{Proof.} A pure-CC realization gives \(q_{K} \left( P\right)=0\) by choosing \(\eta_{A}=\eta_{B}=1.\)
Conversely, if \(q_{K} \left( P\right)=0\), an optimal protocol satisfies
\(\delta=(1-\eta_{A})(1-\eta_{B})=0\), so \(\eta_{A}=1\) or \(\eta_{B}=1\). For \(\eta_{A}=1\), define the
stochastic response
\begin{equation*}
C_{B}^{'} (b|y,\lambda)=\eta_{B} C_{B} (b|y,\lambda)+(1-\eta_{B})Q_{B} (b|y,\lambda).
\end{equation*}
Since Alice uses the same \(C_{A}\) in the CC and CQ branches, exact matching gives
\begin{equation*}
P(a,b|x,y)=\sum_{\lambda=1}^{K} p_{\lambda} C_{A} (a|x,\lambda)C_{B}^{'} (b|y,\lambda).
\end{equation*}
This is a pure-CC realization using the same shared labels. The case \(\eta_{B}=1\) follows by exchanging
Alice and Bob.

\subsection{Sharp eight-label bound for feasible-set saturation}
\label{sec:3.3}

\begin{theorem}
\label{thm:1}
\normalfont
For any two-qubit state and two fixed binary projective measurements at each party, eight shared labels
suffice to preserve the full feasible mode set:
\begin{equation}
\label{eq:15}
\mathcal{F}_{8} \left( P\right)=\mathcal{F}_{\infty} \left( P\right).
\end{equation}
\end{theorem}
Appendix~\ref{sec:A.2} derives the upper bound from the Fenchel--Bunt theorem \cite{Hanner1951}.
Appendices~\ref{sec:A.3}--\ref{sec:A.5} exhibit a physical target and a mode pair that require eight labels.

\subsection{Sharp four-label bound for minimum-cost saturation}
\label{sec:3.4}

\begin{corollary}
\label{cor:1}
\normalfont
Without additional restrictions on the mode probabilities,
\begin{equation}
\label{eq:16}
q_{4} \left( P\right)=q_{\infty} \left( P\right)=\begin{cases}
0, & P\in \mathcal{L}, \\
1, & P\notin \mathcal{L}.
\end{cases}
\end{equation}
\end{corollary}
The classical four-label bound \cite{Donohue2015} and Lemmas~\ref{lem:1}--\ref{lem:2} give
Corollary~\ref{cor:1}. Appendix~\ref{sec:A.6} proves sharpness using the singlet behavior with identical
\(X/Z\) measurements.

\section{Orthogonal measurements of Werner states and resource constraints}
\label{sec:4}

\subsection{Target behavior and measurement conditions}
\label{sec:4.1}

Let \(\zeta\) denote the visibility of the Werner state
\begin{equation}
\label{eq:17}
\rho_{\zeta}=\zeta|\psi^{-} \rangle \langle \psi^{-}|+(1-\zeta)\frac{I_{4}}{4}, 0\le \zeta \le 1,
\end{equation}
with singlet state
\begin{equation*}
|\psi^{-} \rangle = \frac{(|01\rangle -|10\rangle)}{\sqrt{2}}.
\end{equation*}
Each party uses two orthogonal projective measurements:
\begin{equation*}
\mathbf{a}_{0} \cdot \mathbf{a}_{1}=0,\quad \mathbf{b}_{0} \cdot \mathbf{b}_{1}=0.
\end{equation*}
The local means vanish, and the correlators are
\begin{equation}
\label{eq:18}
E_{xy}=\operatorname{Tr} \,[\rho_{\zeta} (A_{x} \otimes B_{y})]=-\zeta \mathbf{a}_{x} \cdot \mathbf{b}_{y}.
\end{equation}
Eq.~\eqref{eq:3} then determines the target behavior.

\subsection{Relative measurement orientations}
\label{sec:4.2}

Let \(\varphi \in [0,\pi]\) be the angle between the oriented unit normals \(\mathbf{a}_{0} \times
\mathbf{a}_{1}\) and \(\mathbf{b}_{0} \times \mathbf{b}_{1}\). Choose a common unit vector in the two
measurement planes as the first axis of each oriented basis. Their overlap matrix is \(\operatorname{diag}
(1,\cos \varphi)\). In the measurement bases this gives
\begin{equation*}
D=R_{A} \operatorname{diag} (1,\cos \varphi)R_{B},\quad E=-\zeta D,
\end{equation*}
where \(D_{xy}=\mathbf{a}_{x} \cdot \mathbf{b}_{y}\) and \(R_{A},R_{B}\) are two-dimensional rotations. Thus
the singular values of \(E\) are \(\zeta\) and \(\zeta|\cos \varphi|\), giving the rank and
smallest-singular-value criteria below.

\subsection{Two-label zero cost and general lower bound}
\label{sec:4.3}

\subsubsection{Rank criterion for zero QQ probability}
\label{sec:4.3.1}

When the local means vanish, Theorem~\ref{thm:3} and Lemma~\ref{lem:2} give
\begin{equation}
\label{eq:19}
q_{2} (P)=0\iff \operatorname{rank} \,E\le 1.
\end{equation}
For the Werner family, this criterion reduces to \(\zeta \cos \,\varphi=0\).

\subsubsection{Lower bound from the smallest singular value}
\label{sec:4.3.2}

\begin{lemma}
\label{lem:3}
\normalfont
Suppose \(P\) is generated by two orthogonal projective measurements per party and has zero local means.
Then
\begin{equation}
\label{eq:20}
q_{2} (P)\ge \frac{\sigma_{\min} (E)}{1+\sigma_{\min} (E)},
\end{equation}
\end{lemma}
where \(\sigma_{\min} (E)\) denotes the smallest singular value of \(E\). The proof is given in
Appendix~\ref{sec:B.1}. For the Werner family,
\begin{equation}
\label{eq:21}
q_{2} (P) \ge \frac{\zeta \left| \cos \,\varphi\right|}{1+\zeta \left| \cos \,\varphi\right|}.
\end{equation}
This bound is positive when \(E\) has full rank.

\section{Complete resource hierarchy for Werner states with identical \texorpdfstring{\(X/Z\)}{X/Z} measurements}
\label{sec:5}

\subsection{Fixed target and exact cost theorem}
\label{sec:5.1}

For identical settings \(A_{0}=B_{0}=X\) and \(A_{1}=B_{1}=Z\), the Werner behavior has \(E=-\zeta I_{2}\) and
\begin{equation}
\label{eq:22}
P_{\zeta} (a,b|x,y)=\frac{1-\zeta ab\delta_{xy}}{4}.
\end{equation}
Here \(\delta_{xy}\) is the Kronecker delta. Write \(q_{K} (\zeta)=q_{K} (P_{\zeta})\) and \(\mathcal{F}_{K}
\left( \zeta\right)=\mathcal{F}_{K} \left( P_{\zeta}\right)\).

\begin{theorem}
\label{thm:2}
\normalfont
For the Werner behavior \(P_{\zeta}\) in Eq.~\eqref{eq:22}, the minimum source cost \(q_K(\zeta)\) is
\begin{equation}
\label{eq:23}
q_{1} (\zeta)=\begin{cases}
0, & \zeta=0, \\
1, & 0<\zeta \le 1.
\end{cases}
\end{equation}

\begin{equation}
\label{eq:24}
q_{2} (\zeta)=\begin{cases}
\frac{\zeta}{1+\zeta}, & 0\le \zeta \le \frac{1}{3}, \\
\left( \sqrt{\frac{2}{1-\zeta^{2}}}-1\right)^{2}, & \frac{1}{3}<\zeta<\frac{1}{\sqrt{2}}, \\
1, & \frac{1}{\sqrt{2}}\le \zeta \le 1.
\end{cases}
\end{equation}

\begin{subequations}
\begin{equation}
\label{eq:25a}
q_{3} (\zeta)=\begin{cases}
0, & 0\le \zeta \le \frac{2}{3}, \\
q_{2} (\zeta), & \frac{2}{3}<\zeta \le 1,
\end{cases}
\end{equation}
and
\begin{equation}
\label{eq:25b}
q_{K} (\zeta)=0\quad (K\ge 4).
\end{equation}
\end{subequations}
The two-label cost is continuous. The three-label cost jumps at \(\zeta=\frac{2}{3}\):
\begin{equation}
\label{eq:26}
q_{3} \left( \frac{2}{3}\right)=0,\quad \lim_{\zeta \downarrow \frac{2}{3}} \,q_{3} (\zeta)=\frac{23-6\sqrt{10}}{5}\approx 0.805.
\end{equation}
\end{theorem}
At the threshold, Eq.~\eqref{eq:D.3} gives a pure-CC protocol with three labels. Above it, the three-label
cost in Fig.~\ref{fig:2} coincides with the two-label curve derived in Appendices~\ref{sec:B} and \ref{sec:C}.
For \(\frac{2}{3}<\zeta<\frac{1}{\sqrt{2}}\), every optimal protocol with at most three shared labels assigns
total weight \(\frac{1}{2}\) to each of two distinct response tuples (Appendix~\ref{sec:D.6}).

\begin{figure}[H]
  \centering
  \includegraphics[width=\linewidth]{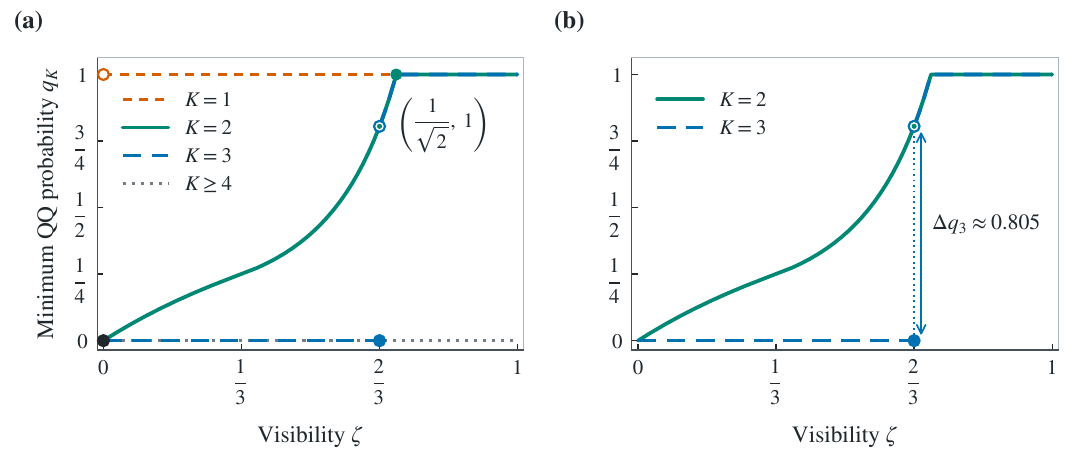}
  \caption{Minimum source cost \(q_K(\zeta)\) (minimum QQ-branch probability) for Werner states with identical
\(X/Z\) measurements. (a) Cost curves by label budget \(K\). (b) Three-label discontinuity at \(\frac{2}{3}\), with the two-label curve
shown for comparison. At \(\zeta=\frac{2}{3}\), the filled blue point gives \(q_3=0\). The blue ring marks the
right-hand limit of \(q_3\), and its green center marks the attained value of \(q_2\).}
  \label{fig:2}
\end{figure}

Appendix~\ref{sec:D.2} bounds the cost away from zero above this threshold. Appendix~\ref{sec:D.5} derives the
right-hand limit in Eq.~\eqref{eq:26} using the global cover in Appendix~\ref{sec:D.3} and the row bounds in
Appendix~\ref{sec:D.4}, both certified in Ref.~\cite{Verification31}.

The worst-case total variation distance between \(P_\zeta\) and \(P_{\frac{2}{3}}\) is \(\tfrac{1}{2}\left|\zeta-\tfrac{2}{3}\right|\), as follows from Eq.~\eqref{eq:22}. The three-label protocol at the threshold therefore
approximates \(P_\zeta\) with zero source cost and error at most \(\varepsilon>0\) throughout
\begin{equation*}
\frac{2}{3}<\zeta \le \min \{1,\frac{2}{3}+2\varepsilon \}.
\end{equation*}

\subsection{Complete four-label feasible mode set}
\label{sec:5.2}

For the Werner behaviors \(P_\zeta\) of Eq.~\eqref{eq:22}, four shared labels realize every feasible mode pair, so
\(\mathcal{F}_{4} (\zeta)=\mathcal{F}_{\infty} (\zeta)\). For \(0\le \zeta \le \frac{1}{\sqrt{2}}\), this set
is \([0,1]^{2}\). For \(\frac{1}{\sqrt{2}}<\zeta \le 1\), define
\begin{equation}
\label{eq:27}
r_{\zeta}=\frac{\sqrt{2}\zeta-1}{\sqrt{2}-1}.
\end{equation}
The feasible set consists of the pure-QQ point \((0,0)\) and the region
\begin{equation}
\label{eq:28}
r_{\zeta} \le \eta_{A} \le 1, \frac{r_{\zeta} \eta_{A}}{(1+r_{\zeta})\eta_{A}-r_{\zeta}}\le \eta_{B} \le 1.
\end{equation}
The denominator is positive throughout this region. At \(\zeta=1\), only pure CC and pure QQ are feasible.
Appendix~\ref{sec:E.1} proves this characterization.

\subsection{Thresholds for cost and feasible-set saturation}
\label{sec:5.3}

\begin{figure}[!htbp]
  \centering
  \includegraphics[width=\linewidth]{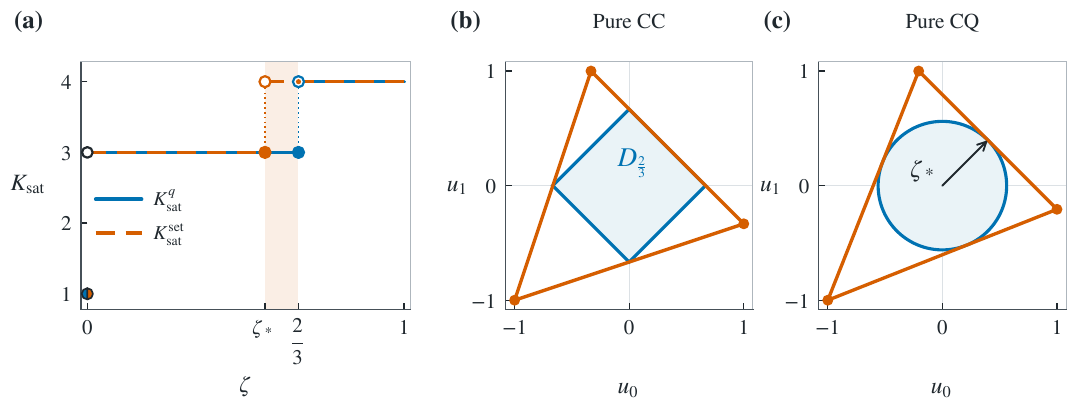}
  \caption{Shared-label saturation for Werner states with identical \(X/Z\) measurements. (a) Capacities for minimum-cost
and feasible-set saturation. Shading marks the interval where the two capacities differ. Filled and open circles denote
included and excluded endpoints. At \(\zeta=\frac{2}{3}\), the blue point at three and the orange dot inside the
blue ring at four give the cost and feasible-set capacities, respectively. (b) Triangle enclosing the diamond \(D_{\frac{2}{3}}=\{x:\Vert x\Vert_{1} \le
\frac{2}{3}\}\), attaining the three-label pure-CC threshold. The triangle in Eq.~\eqref{eq:D.3} is reflected across the \(u_{0}\) axis here. (c) Triangle enclosing a centered disk of radius
\(\zeta_{\ast} \approx 0.5609\), attaining the three-label pure-CQ threshold.}
  \label{fig:3}
\end{figure}

Define the label capacities for cost saturation and feasible-set saturation by
\begin{align}
\label{eq:29}
K_{\mathrm{sat}}^{q} (\zeta)&=\min \{K:q_{K} (\zeta)=q_{\infty} (\zeta)\},\\
\label{eq:30}
K_{\mathrm{sat}}^{\mathrm{set}} (\zeta)&=\min \{K:\mathcal{F}_{K} (\zeta)=\mathcal{F}_{\infty} (\zeta)\}.
\end{align}
Figure \ref{fig:3}(a) gives the exact capacities, including the threshold endpoints. The pure-CQ boundary has
the following equivalent measurement interpretation.

\begin{proposition}[Sharp three-outcome threshold]
\label{prop:1}
\normalfont
The Werner behavior with identical \(X/Z\) measurements admits a pure-CQ realization with at most three shared
labels exactly for \(0\le \zeta \le \zeta_{\ast}\). This is also the visibility range in which the noisy Pauli
measurements \(M_{a|x}=\frac{I+a\zeta A_{x}}{2}\) can be obtained by stochastic postprocessing of a parent
measurement with at most three outcomes. The threshold \(\zeta_{\ast}\) is the unique positive solution of
\begin{equation*}
\zeta_{\ast}^{6}+4\zeta_{\ast}^{4}+5\zeta_{\ast}^{2}=2,\quad \zeta_{\ast} \approx 0.5609.
\end{equation*}
\end{proposition}
Reference \cite{Skrzypczyk2020} gives a three-outcome construction in its Eqs.~(32)--(33) and leaves
optimality open. Appendix~\ref{sec:E.2} establishes the correspondence, and Appendix~\ref{sec:E.3} proves the matching upper bound.

Das et al.~\cite{Das2018} reported a four-label minimum for LHV--LHS simulation of noisy BB84 correlations throughout
\(0<\zeta\le\frac{1}{\sqrt{2}}\). Flipping Alice's output for \(x=0\) maps their noisy BB84 behavior
to the Werner behavior \(P_\zeta\) in Eq.~\eqref{eq:22}.
For example, at \(\zeta=\frac{4}{9}\), take \(p_j,u_j\) from Eq.~\eqref{eq:D.3} and set \(s_j=-\frac{2}{3}u_j\).
Both means vanish, \(\max_j\|s_j\|^2=\frac{8}{9}\), and \(\overline{us^{\mathsf T}}=-\frac{4}{9}I_2\).
These three labels thus realize the behavior with stochastic responses for Alice and qubit states for Bob.

For pure CQ, one label suffices at \(\zeta=0\). The rank bound rules out two labels at positive visibility, so
the minimum is three for \(0<\zeta \le \zeta_{\ast}.\) It rises to four for \(\zeta_{\ast}<\zeta \le
\frac{1}{\sqrt{2}}\). Above \(\frac{1}{\sqrt{2}}\), no pure-CQ realization exists. Appendices~\ref{sec:E.4}
and~\ref{sec:E.5} determine the feasible-set capacity for general mode pairs.

For \(\zeta_{\ast}<\zeta\le\frac{2}{3}\),
\begin{equation*}
K_{\mathrm{sat}}^{q} (\zeta)=3,\quad K_{\mathrm{sat}}^{\mathrm{set}} (\zeta)=4.
\end{equation*}
In this interval, three labels attain zero source cost in pure CC, and the pure-CQ mode requires a fourth label.

\section{Local marginals and source cost at fixed correlation tensor}
\label{sec:6}

\subsection{Centered-correlation criterion for two-label zero cost}
\label{sec:6.1}

\begin{theorem}
\label{thm:3}
\normalfont
Let \(P\) be a nonsignaling behavior with finite input sets and binary outputs. Define its centered
correlation matrix by
\begin{equation}
\label{eq:31}
\Gamma=E-m_{A} {m_{B}}^{\mathsf{T}}.
\end{equation}
Then \(P\) admits a classical realization with stochastic local responses and at most two shared labels
if and only if \(\operatorname{rank} \,\Gamma \le 1\).
\end{theorem}

Under the assumptions of Lemma~\ref{lem:2}, this gives
\begin{equation}
\label{eq:32}
q_{2} (P)=0\iff \operatorname{rank} \,\Gamma \le 1.
\end{equation}
The probability matrix \(M_{(x,a),\allowbreak(y,b)}=P(a,b|x,y)\) has \(\operatorname{rank}
1+\operatorname{rank} \Gamma.\) When this rank is at most two, it equals the nonnegative rank
\cite{Cohen1993}. Appendix~\ref{sec:F.1} explicitly constructs normalized local responses and a common label
distribution for all input pairs.

\subsection{Source-cost separation at fixed correlation tensor}
\label{sec:6.2}

To examine the role of local marginals, consider the two-qubit state family
\begin{equation}
\label{eq:33}
\rho_{\zeta,\mu}=\frac{1}{4}[I\otimes I+\mu (Z\otimes I-I\otimes Z)-\zeta (X\otimes X+Y\otimes Y+Z\otimes Z)].
\end{equation}
Both parties use fixed measurements \(A_{0}=B_{0}=X\) and \(A_{1}=B_{1}=Z\). The parameters satisfy
\begin{equation}
\label{eq:34}
0<\zeta \le \frac{1}{3},|\mu|\le t_{\max} \left( \zeta\right),\quad t_{\max} \left( \zeta\right)=\frac{1}{2}\sqrt{\left( 1-\zeta\right) \left( 1+3\zeta\right)}.
\end{equation}
Appendix~\ref{sec:F.2} verifies separability throughout this range using the positive-partial-transpose
criterion \cite{Peres1996,Horodecki1996}. The measurements give
\begin{equation}
\label{eq:35}
m_{A}=-m_{B}=(0,\mu)^{\mathsf{T}},E=-\zeta I_{2}.
\end{equation}
The full correlation tensor is defined by \(T_{ij}=\operatorname{Tr} \,[\rho (\sigma_{i} \otimes
\sigma_{j})]\), where \(i,j\in \{x,y,z\}\). Then
\begin{equation}
\label{eq:36}
T=-\zeta I_{3},\Gamma=\operatorname{diag} (-\zeta,\mu^{2}-\zeta).
\end{equation}
At fixed \(\zeta\), varying \(\mu\) changes the local means while leaving \(E\) and \(T\) fixed.
Conjugation by \(X\otimes X\) sends \(\rho_{\zeta,\mu}\) to \(\rho_{\zeta,-\mu}\) and reverses both \(Z\) outcomes.
The same relabeling of the classical responses preserves each feasible mode pair. Thus the two-label cost
\(q_{2}(\zeta,\mu)\) depends on \(\mu\) only through the local bias \(t=|\mu|\).
By Eqs.~\eqref{eq:32} and~\eqref{eq:36},
\begin{equation}
\label{eq:37}
q_{2} (\zeta,\mu)=0\iff \mu^{2}=\zeta.
\end{equation}
Together with the Werner result at \(\mu=0\), this gives
\begin{equation}
\label{eq:38}
q_{2} (\zeta,0)=\frac{\zeta}{1+\zeta},q_{2} \left( \zeta,\sqrt{\zeta}\right)=0.
\end{equation}
The cost difference is \(\frac{1}{4}\) at \(\zeta=\frac{1}{3}.\) Appendix~\ref{sec:F.2} constructs an explicit
zero-cost protocol.

\begin{figure}[H]
  \centering
  \includegraphics[width=\linewidth]{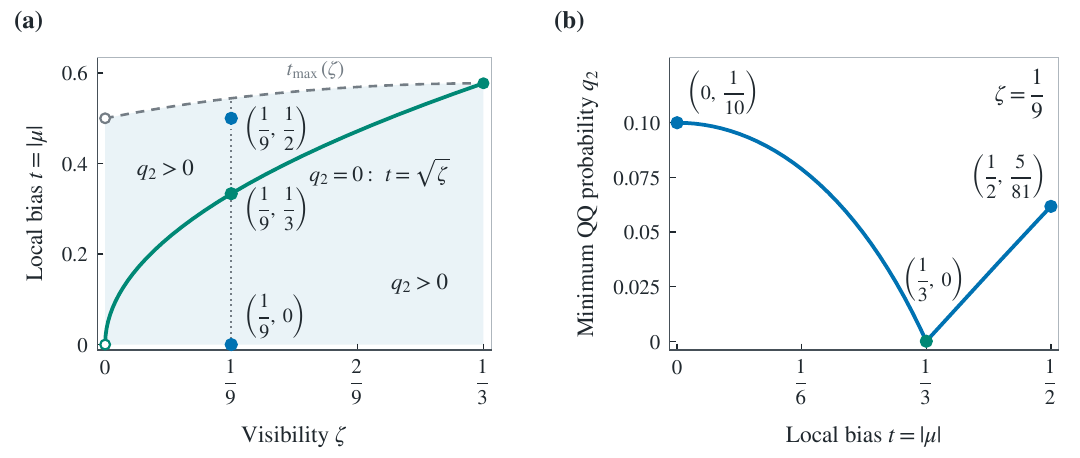}
  \caption{Two-label cost \(q_2\) versus local bias \(t=|\mu|\). (a) Shading marks the range in
Eq.~\eqref{eq:34} for the separable family in Eq.~\eqref{eq:33}. The zero-cost curve is \(t=\sqrt{\zeta}\)
and the dotted line marks \(\zeta=\frac{1}{9}\). (b) Exact cost on this slice for \(0\le t\le
\frac{1}{2}\), \(E=-\frac{I_{2}}{9}\) and \(T=-\frac{I_{3}}{9}\) held fixed.}
  \label{fig:4}
\end{figure}

\subsection{Nonmonotonic dependence on local bias}
\label{sec:6.3}

For fixed \(0<\zeta<\frac{1}{3}\), the two-label cost satisfies
\begin{subequations}
\begin{equation}
\label{eq:39a}
q_{2} (\zeta,0)>0,q_{2} \left( \zeta,\sqrt{\zeta}\right)=0,q_{2} (\zeta,t_{\max})>0.
\end{equation}
Since \(0<\sqrt{\zeta}<t_{\max}\), Eq.~\eqref{eq:39a}, derived in Appendix~\ref{sec:F.2}, implies that the cost is nonmonotonic in \(t=|\mu|\). At \(\zeta=\frac{1}{9}\), Fig.~\ref{fig:4}(b) shows the exact curve
\begin{equation}
\label{eq:39b}
q_{2} \left( \frac{1}{9},t\right)=\begin{cases}
1-\frac{9}{5+\sqrt{25-81t^{2}}}, & 0\le t\le \frac{1}{3}, \\
\frac{9t^{2}-1}{9(1+t)^{2}}, & \frac{1}{3}<t\le \frac{1}{2}.
\end{cases}
\end{equation}
\end{subequations}
Appendix~\ref{sec:F.3} proves Eq.~\eqref{eq:39b} and gives attaining protocols. The curve decreases to zero
and then increases, with a cusp at \(t=\frac{1}{3}\).

\section{Distributed sampling with local source queries}
\label{sec:7}

Distributed sampling asks what resources two parties need to generate prescribed correlations
\cite{Toner2003,Degorre2005}. Under dimension constraints, quantum protocols can have an advantage in
generating shared randomness \cite{Ma2023}. Here the target law is defined by two independent doubly symmetric
binary sources \(\left( U_{i},V_{i}\right),i\in \{0,1\}\). Each \(U_{i}\) is uniform on \(\{-1,+1\}\), and
\(V_{i}\) is obtained by flipping its sign with probability \(p\in \left[ 0,\frac{1}{2}\right]\).

The parties use the interfaces of Sec.~\ref{sec:2} with at most \(K\) shared labels and free private
randomness. After resource preparation, the query pair \((x,y)\in \{0,1\}^{2}\) is drawn independently of the
resources. The task is to reproduce the joint law of \((U_{x},V_{y})\) without communication, with Alice
producing \(u\) on input \(x\) and Bob producing \(v\) on input \(y\).
\begin{equation}
\label{eq:40}
P_{p}^{\mathrm{samp}} (u,v|x,y)=\frac{1}{4}\left( 1+(1-2p)uv\delta_{xy}\right).
\end{equation}
With \(\zeta=1-2p\), Bob's output flip \((u,v)=(a,-b)\) maps the Werner behavior \(P_\zeta\)
in Eq.~\eqref{eq:22} to the sampling target \(P_p^{\mathrm{samp}}\) in Eq.~\eqref{eq:40}.
Theorem~\ref{thm:2} therefore gives the exact minimum source cost for each label budget. In particular,
\begin{equation}
\label{eq:41}
q_{2} \left( P_{p}^{\mathrm{samp}}\right)=\frac{1-2p}{2-2p},\quad q_{3} \left( P_{p}^{\mathrm{samp}}\right)=0,\frac{1}{3}\le p\le \frac{1}{2}.
\end{equation}
At \(p=\frac{1}{3}\), the prescribed Werner state \(\rho_{\frac{1}{3}}\) is separable, and an optimal
two-label protocol attains \((\delta,N_{\mathrm{Q}})=(\frac{1}{4},1)\) (Appendix~\ref{sec:G.2}). With the same label
budget, the CC/QQ baseline requires \((\delta,N_{\mathrm{Q}})=(1,2)\) even with coordinated interface choices
(Appendix~\ref{sec:G.1}). Allowing CQ and QC reduces mean source calls by 75\% and mean quantum-interface
activations by 50\% relative to this baseline.

At zero source cost, protocols with at most two shared labels incur a positive sampling error for
\(0\le p<\frac{1}{2}\). Define the worst-case total variation error by
\begin{equation*}
d(P,Q)=\max_{x,y} \,\frac{1}{2}\sum_{a,b}|P(a,b|x,y)-Q(a,b|x,y)|.
\end{equation*}
Let \(\mathcal{C}_{2}\) denote the pure-CC behaviors realizable with at most two shared labels. Then
\begin{equation}
\label{eq:42}
\min_{Q\in \mathcal{C}_{2}} \,d(Q,P_{p}^{\mathrm{samp}})=\sqrt{\frac{3}{2}-p}-1,\;0\le p\le \frac{1}{2}.
\end{equation}
By Lemma~\ref{lem:2}, protocols with zero source cost and at most two shared labels generate exactly the
behaviors in \(\mathcal{C}_{2}\). Eq.~\eqref{eq:42} therefore gives the minimum error tolerance for sampling
without the source. At \(p=\frac{1}{3}\), this tolerance is approximately \(0.08\). Appendix~\ref{sec:G.3}
proves the lower bound and gives an attaining protocol.

\section{Conclusions}
\label{sec:8}

In this hybrid model, each party chooses its interface independently. The classical interface uses the shared label.
In the CQ and QC branches, the quantum interface measures a label-dependent qubit state.
Only the QQ branch uses the prescribed two-qubit source. Four shared labels
suffice to attain the minimum source cost for any target. Preserving all feasible pairs of interface choice
probabilities can require eight. Within a family of separable states, the two-label cost varies nonmonotonically
with local bias even though the full correlation tensor stays fixed. Werner states exhibit a different effect
under identical \(X\) and \(Z\) measurements: the three-label cost jumps from zero as the visibility rises
above \(\dfrac{2}{3}\). The Werner cost hierarchy determines the exact source cost of distributed sampling at each
label budget. For classical sampling with two shared labels, the analysis gives the minimum worst-case total
variation error at each flip probability.

These costs are defined for the prescribed source and independent interface controllers. Determining the full
cost under approximate simulation and for general measurement orientations remains open. A further direction
is to study coordinated interface choices while counting any shared control randomness in the resource budget.

\section*{\phantomsection Acknowledgments}
\addcontentsline{toc}{section}{Acknowledgments}

The authors used ChatGPT and Codex (OpenAI) to assist with literature searches, manuscript editing and
preparation of the accompanying code. The authors checked the references and verified the numerical results
against the analytical expressions in the paper. All authors take responsibility for the scientific content.

\paragraph{Funding information}
This work is supported by the National Natural Science Foundation of China
(Grants 12475020 and 92565111), Quantum Science and Technology-National
Science and Technology Major Project (2021ZD0301701), and the National Key
Research and Development Program of China (2023YFC2205802).

\section*{\phantomsection Data availability}
\label{sec:data}
\addcontentsline{toc}{section}{Data availability}

Appendices~\ref{sec:A}--\ref{sec:G} contain the analytic derivations. The version 31.2 archive
\cite{Verification31} provides the exact certificates, code, instructions and reports for the finite checks in
Appendix~\ref{sec:D}. Its version manifest records an earlier manuscript revision. The
\href{https://github.com/Chenzhnju/hybridization-analysis/blob/v31.2/docs/MANUSCRIPT_MAP.md}{verification guide}
matches the checks to the appendix sections, equations and tables used here.

\appendix
\numberwithin{equation}{section}
\renewcommand{\theHequation}{appendix.\Alph{section}.\arabic{equation}}
\section{General structure and sharp uniform label bounds}
\label{sec:A}

\subsection{Model and notation}
\label{sec:A.1}

Use the eight moments \(z(P)=({m_{A}}^{\mathsf{T}},{m_{B}}^{\mathsf{T}},\left( \operatorname{vec}
\,E\right)^{\mathsf{T}})^{\mathsf{T}}\) from Sec.~\ref{sec:2}. Let \(\mathcal{Q}_{A}\) and \(\mathcal{Q}_{B}\)
denote Alice's and Bob's compact, convex quantum-response sets. For \(u,v\in [-1,1]^{2}\), \(r\in
\mathcal{Q}_{A}\), and \(s\in \mathcal{Q}_{B}\), the single-label contribution is
\begin{equation}
\label{eq:A.1}
g(u,v,r,s)=\begin{pmatrix}
\eta_{A} u+\gamma r \\
\eta_{B} v+\beta s \\
\operatorname{vec} \,(wuv^{\mathsf{T}}+\beta us^{\mathsf{T}}+\gamma rv^{\mathsf{T}})
\end{pmatrix}.
\end{equation}
At fixed mode probabilities, a protocol with at most \(K\) shared labels is feasible if and only if
\begin{equation}
\label{eq:A.2}
\sum_{j=1}^{K} p_{j}g\left( u_{j},v_{j},r_{j},s_{j}\right)=\tau z\left( P\right),\quad p_{j} \ge 0, \sum_{j=1}^{K} p_{j}=1.
\end{equation}
For fixed \(K\), Eq.~\eqref{eq:A.2} defines a closed subset of the compact parameter space. Since pure QQ is
feasible, the continuous cost in Eq.~\eqref{eq:12} attains its minimum.

\subsection{Uniform eight-label upper bound}
\label{sec:A.2}

The Fenchel--Bunt theorem~\cite{Hanner1951} states that every point in the convex hull of a connected set \(S\) of affine dimension \(d\geq 1\) can be expressed as a convex combination of at most \(d\) points of \(S\).

The parameter space \(\mathcal{D}=[-1,1]^{2} \times [-1,1]^{2} \times \mathcal{Q}_{A} \times \mathcal{Q}_{B}\)
is compact and convex, so its continuous image \(\mathcal{S}=g(\mathcal{D})\subseteq \mathbb{R}^{8}\) is
compact and path connected. Applying the theorem to the residual \(\tau z(P)\in \operatorname{conv}
\mathcal{S}\) in Eq.~\eqref{eq:A.2} gives at most eight admissible response tuples at the same mode pair. For
\(\tau=0\), operation is pure QQ and labels are irrelevant. Hence \(\mathcal{F}_{8} (P)=\mathcal{F}_{\infty}
(P)\).

\subsection{An exposed face with exactly eight single-label points}
\label{sec:A.3}

To prove sharpness, choose \(A_{0}=B_{0}=X\) and \(A_{1}=B_{1}=Z\) with the quantum-response sets specified by
Eq.~\eqref{eq:7}. Set \(\eta_{A}=\eta_{B}=\frac{1}{2}\). The branch weights are
\begin{equation*}
w=\beta=\gamma=\delta=\frac{1}{4},\tau=\frac{3}{4}.
\end{equation*}
Define \(H=\begin{pmatrix}
1 & 1 \\
1 &-1
\end{pmatrix},W(z)=E_{00}+E_{01}+E_{10}-E_{11}.\) For any admissible single-label contribution \(g\),

\begin{equation}
\label{eq:A.3}
W(g)=\frac{1}{4}(u^{\mathsf{T}} Hv+u^{\mathsf{T}} Hs+r^{\mathsf{T}} Hv)\le \frac{3}{2}.
\end{equation}
Maximizing the first term over the corners of the two response squares gives \(u^{\mathsf{T}}Hv\le2\). Since \(H^{\mathsf{T}}H=2I_{2}\) and \(r,s\) lie in the unit disk, the other two terms are also bounded by 2.
Equality in Eq.~\eqref{eq:A.3} requires \(\Vert u\Vert=\Vert v\Vert=\sqrt{2}\). All three bounds are saturated
simultaneously if and only if
\begin{equation}
\label{eq:A.4}
u,v\in \{-1,+1\}^{2},\quad u^{\mathsf{T}} Hv=2,r=\frac{Hv}{2},s=\frac{H^{\mathsf{T}} u}{2}.
\end{equation}
Eq.~\eqref{eq:A.4} fixes \(v_{0}=u_{0}\) if \(u_{0}=u_{1}\) and \(v_{1}=u_{0}\) otherwise. Take \(u\) in the
order \((-1,-1),\allowbreak(-1,+1),\allowbreak(+1,-1),\allowbreak(+1,+1)\), with the free coordinate of \(v\)
taking \(-1\) and then \(+1\) for each \(u\). Write \(g_{1},\ldots,g_{8}\) for the resulting contributions.
Then
\begin{equation}
\label{eq:A.5}
\mathcal{S}\cap \{W=\frac{3}{2}\}=\{g_{1},\ldots,g_{8} \}.
\end{equation}

\subsection{Physical target from a rotated Werner state}
\label{sec:A.4}

Consider the Werner state \(\rho_{\frac{1}{\sqrt{2}}}\) in Eq.~\eqref{eq:17}. Applying a local unitary to
Bob's subsystem gives the target state
\begin{equation}
\label{eq:A.6}
\rho_{\ast}=(I\otimes U)\rho_{\frac{1}{\sqrt{2}}} (I\otimes U^{\dagger}),\quad U=\cos \,\left( \frac{\pi}{8}\right) Z-\sin \,\left( \frac{\pi}{8}\right) X.
\end{equation}
Under the fixed \(X/Z\) measurements, the resulting behavior \(P_{\ast}\) has moments
\begin{equation}
\label{eq:A.7}
m_{A}=m_{B}=0,E_{\ast}=\frac{1}{2}H.
\end{equation}
Give equal weight to the eight response pairs \((u,v)\) specified by Eq.~\eqref{eq:A.4}. Then \(u\) and \(v\)
each have zero mean and second-moment matrix \(I_{2}\),
and \(\mathbb{E}[v\mid u]=\frac{Hu}{2}\). Eq.~\eqref{eq:A.1} gives
\begin{equation}
\label{eq:A.8}
\frac{1}{8}\sum_{j=1}^{8} g_{j}=\tau z\left( P_{\ast}\right).
\end{equation}

\subsection{Affine independence and eight-label necessity}
\label{sec:A.5}

Let \(N\) have rows \(g_{j}-g_{8}\), \(1\le j\le 7\), restricted in order to the four local means and the
\(00,01,10\) correlation coordinates. The ordering in Appendix~\ref{sec:A.3} and Eq.~\eqref{eq:A.1}
give
\begin{equation}
\label{eq:A.9}
\det N=-\frac{49}{16}.
\end{equation}
The eight points are therefore affinely independent. Since the residual saturates the bound in
Eq.~\eqref{eq:A.3}, every positive-weight term equals one of the eight points in Eq.~\eqref{eq:A.5}. The uniform
decomposition in Eq.~\eqref{eq:A.8} is unique, so
\begin{equation}
\label{eq:A.10}
\left( \frac{1}{2}, \frac{1}{2}\right) \in \mathcal{F}_{8} \left( P_{\ast}\right),\left( \frac{1}{2},\frac{1}{2}\right) \notin \mathcal{F}_{7} \left( P_{\ast}\right).
\end{equation}
Together with the upper bound in Appendix~\ref{sec:A.2}, this proves Theorem~\ref{thm:1}.

\subsection{Sharp four-label bound for minimum-cost saturation}
\label{sec:A.6}

Lemma~\ref{lem:1} gives \(q_{K} (P)=1\) for every nonlocal target. If \(P\) is local, decompose Alice's
responses into the four deterministic strategies \(u\in \{-1,1\}^{2}\) and average Bob's responses conditional
on \(u\) \cite{Donohue2015}. The resulting pure-CC protocol uses at most four labels, so \(q_{4} (P)=0\).

For the singlet with identical \(X/Z\) measurements, \(m_{A}=m_{B}=0\) and \(E=-I_{2}\). In a pure-CC
realization, perfect anticorrelation requires \(u\in \{-1,1\}^{2}\) and \(v=-u\) for every positive-weight
label. The remaining moment constraints are
\begin{equation*}
\mathbb{E}[u_{0}]=\mathbb{E}[u_{1}]=\mathbb{E}[u_{0} u_{1}]=0.
\end{equation*}
Each of the four strategies therefore has weight \(\tfrac{1}{4}\). By Lemma~\ref{lem:2},
\(q_{3} (P)>q_{4} (P)=0\).

\section{Two-label bounds and attaining protocols}
\label{sec:B}

\subsection{Rank constraints and source-cost lower bound}
\label{sec:B.1}

For zero local means, Theorem~\ref{thm:3} gives the criterion \(\operatorname{rank} E\le 1\), and Eq.~\eqref{eq:19} then follows from Lemma~\ref{lem:2}. The necessary direction is the specialization of
Eq.~\eqref{eq:F.1}:
\begin{equation}
\label{eq:B.1}
E=p_{1} p_{2} (u_{1}-u_{2})(v_{1}-v_{2})^{\mathsf{T}}.
\end{equation}

For targets with zero local means considered in Lemma~\ref{lem:3}, orthogonal local measurements imply \(\Vert
r_{\lambda} \Vert,\Vert s_{\lambda} \Vert \le 1\). Define the effective response vectors
\begin{equation}
\label{eq:B.2}
\xi_{\lambda}=\eta_{A} u_{\lambda}+(1-\eta_{A})r_{\lambda},\quad \chi_{\lambda}=\eta_{B} v_{\lambda}+(1-\eta_{B})s_{\lambda}.
\end{equation}
The moment-matching conditions give
\begin{equation}
\label{eq:B.3}
\overline{\xi}=\delta \overline{r}, \overline{\chi}=\delta \overline{s},\quad \overline{\xi \chi^{\mathsf{T}}}=\tau E+\delta \overline{rs^{\mathsf{T}}}.
\end{equation}
Write \(H=\operatorname{Cov} (\xi,\chi)\). With at most two labels, \(H\) has rank at most one.
Eq.~\eqref{eq:B.3} gives
\begin{equation}
\label{eq:B.4}
H=\tau E+\delta D,\quad D=\overline{rs^{\mathsf{T}}}-\delta \overline{r}\overline{s}^{\mathsf{T}}.
\end{equation}
The form \(\langle f,g\rangle_{\delta}=\mathbb{E}[fg]-\delta \mathbb{E}[f]\mathbb{E}[g]\) is positive
semidefinite. Applied to scalar projections of \(r,s\), the Cauchy--Schwarz inequality and the unit-disk
bounds give \(\Vert D\Vert_{\mathrm{op}} \le 1\). On a unit right null vector of \(H\), Eq.~\eqref{eq:B.4}
yields \(\tau \sigma_{\min} (E)\le \delta\), hence \(q_{2} (P)\ge \frac{\sigma_{\min} (E)}{1+\sigma_{\min}
(E)}\).

For one label, \(H=0\) and \(D=\tau rs^{\mathsf{T}}\). If \(\delta<1\), then \(E=-\delta rs^{\mathsf{T}}\),
excluding \(E=-\zeta I_{2}\) for \(\zeta>0\). Pure QQ attains unit cost, while one fair, independent classical
output per party suffices at \(\zeta=0\). This proves Eq.~\eqref{eq:23}.

\subsection{Two-label protocols attaining the lower bounds}
\label{sec:B.2}

Take two equiprobable labels, \(\eta_{A}=\eta_{B}=1-d\) with \(0<d<1\), and set \(a=\min \{d,1-d\}\) and
\(b=\sqrt{2d^{2}-a^{2}}\). For one label choose
\begin{equation}
\label{eq:B.5}
\begin{gathered}
u=\frac{a}{1-d}(1,1)^{\mathsf{T}},\quad v=\frac{a}{1-d}(1,-1)^{\mathsf{T}}, \\
r=\frac{1}{2d}(-a-b,b-a)^{\mathsf{T}},\quad s=\frac{1}{2d}(-a-b,a-b)^{\mathsf{T}}.
\end{gathered}
\end{equation}
Reverse all four vectors for the other label. The local means vanish, the classical responses lie in the
square, and \(a^{2}+b^{2}=2d^{2}\) places both quantum responses on the unit circle. Substitution gives
\(wuv^{\mathsf{T}}+\beta us^{\mathsf{T}}+\gamma rv^{\mathsf{T}}=-abI_{2}\). Thus the cost is \(\delta=d^{2}\)
and the visibility is
\begin{equation}
\label{eq:B.6}
\zeta=
\begin{cases}
\dfrac{d^{2}}{1-d^{2}}, & 0<d\le \frac12,\\
Z(d), & \frac12<d<1,
\end{cases}
\qquad Z(d)=\sqrt{1-\frac{2}{(1+d)^2}}.
\end{equation}
The protocol attains the singular-value bound for \(0<\zeta\le\frac{1}{3}\) and the bound of
Appendix~\ref{sec:C} for \(\frac{1}{3}<\zeta<\frac{1}{\sqrt{2}}\). At \(\zeta=0\) one pure-CC label suffices.
For \(\zeta \ge \frac{1}{\sqrt{2}}\), Appendix~\ref{sec:C.3} rules out \(\delta<1\) with at most two labels,
and pure QQ attains unit cost.

\section{Higher-visibility two-label lower bound}
\label{sec:C}

\subsection{Moment reduction}
\label{sec:C.1}

Consider a two-label protocol for \(P_{\zeta}\) with \(\zeta>\frac{1}{3}\) and \(\delta<1\).
Appendix~\ref{sec:B.1} gives \(\delta>\frac14\). Thus \(d=\sqrt{\delta}\) belongs to \((\frac12,1)\) and \(\tau>0\). If \(\eta_{A} \eta_{B}=0\), the only non-QQ branch would
reproduce \(P_{\zeta}\) as a mixture of at most two product behaviors. Eq.~\eqref{eq:B.1} would then give
\(\operatorname{rank} (E)\le 1\), whereas \(E=-\zeta I_{2}\) has rank two. Thus \(0<\eta_{A},\eta_{B}<1\).
Exchanging the parties replaces the scalar \(s=\frac{d\eta_{A}}{\gamma}\) by its reciprocal, so take \(0<s\le 1\). Dividing the non-QQ branch weights by \(\tau\) gives
\begin{equation*}
\tilde{w}=\frac{s(1-d^{2})}{(d+s)(1+ds)},\tilde{\beta}=\frac{ds}{1+ds},\tilde{\gamma}=\frac{d}{d+s}.
\end{equation*}
The normalized weights sum to one. With \(\Delta f=\sqrt{p_{1} p_{2}}(f_{1}-f_{2})\), the two-label covariance
identity reads
\begin{equation*}
\overline{fg^{\mathsf{T}}}=\overline{f}\overline{g}^{\mathsf{T}}+\Delta f(\Delta g)^{\mathsf{T}}.
\end{equation*}
Here \(\overline{s}\) and \(\Delta s\) denote the mean and weighted difference of Bob's response vectors \(s_{\lambda}\).
The zero target local means imply \(\overline{r}=-\frac{s}{d}\overline{u}\) and
\(\overline{s}=-\frac{1}{ds}\overline{v}\). Substituting into the matching conditions gives
\begin{equation}
\label{eq:C.1}
\begin{gathered}
-\zeta I_{2}=-\overline{u}\overline{v}^{\mathsf{T}}+\tilde{w}\Delta u(\Delta v)^{\mathsf{T}}+\tilde{\beta}\Delta u(\Delta s)^{\mathsf{T}}+\tilde{\gamma}\Delta r(\Delta v)^{\mathsf{T}}, \\
\Vert \overline{u}\Vert^{2}+\Vert \Delta u\Vert^{2} \le 2,\frac{s^{2}}{d^{2}}\Vert \overline{u}\Vert^{2}+\Vert \Delta r\Vert^{2} \le 1, \\
\quad \Vert \overline{v}\Vert^{2}+\Vert \Delta v\Vert^{2} \le 2,\frac{1}{d^{2} s^{2}}\Vert \overline{v}\Vert^{2}+\Vert \Delta s\Vert^{2} \le 1.
\end{gathered}
\end{equation}
The norm bounds follow by averaging the classical and quantum response constraints.

\subsection{Weighted spectral bound}
\label{sec:C.2}

\textit{Spectral estimate.} Let \(D\in \mathbb{R}^{3\times 3}\) be invertible, with singular values
\(\sigma_{1} \ge \sigma_{2} \ge \sigma_{3}>0\). If \(X,Y\in \mathbb{R}^{2\times 3}\) satisfy
\(XDY^{\mathsf{T}}=\pm zI_{2}\) with \(z>0\), then
\begin{equation}
\label{eq:C.2}
\Vert X\Vert_{\mathrm{F}} \Vert Y\Vert_{\mathrm{F}} \ge z\left( \frac{1}{\sigma_{1}}+\frac{1}{\sigma_{2}}\right).
\end{equation}

\textit{Proof.} Let \(t_{1} \ge t_{2}>0\) denote the singular values of \(XD.\) The minimum-norm solution of
\(XDY^{\mathsf{T}}=\pm zI_{2}\) gives \(\Vert Y\Vert_{\mathrm{F}}^{2} \ge z^{2} (t_{1}^{-2}+t_{2}^{-2}).\) Since
\(X=(XD)D^{-1}\), the singular-value rearrangement gives \(\Vert X\Vert_{\mathrm{F}}^{2} \ge
\frac{t_{1}^{2}}{\sigma_{1}^{2}}+\frac{t_{2}^{2}}{\sigma_{2}^{2}}\). Multiplying these bounds and applying the Cauchy--Schwarz inequality gives Eq.~\eqref{eq:C.2}. To apply this estimate, choose weights \(a,b>0\) and form the following \(2\times3\) matrices.
\begin{equation*}
X=\left( \frac{\sqrt{ad^{2}+s^{2}}}{d}\overline{u},\sqrt{a}\Delta u,\Delta r\right),\quad Y=\left( \frac{\sqrt{bd^{2} s^{2}+1}}{ds}\overline{v},\sqrt{b}\Delta v,\Delta s\right).
\end{equation*}
These weights combine the classical and quantum norm constraints. Indeed, Eq.~\eqref{eq:C.1} gives \(\Vert
X\Vert_{\mathrm{F}}^{2} \le\) \(1+2a\), \(\Vert Y\Vert_{\mathrm{F}}^{2} \le 1+2b\). Define \(\sigma_{0}=\frac{d^{2}
s}{\sqrt{(ad^{2}+s^{2})(bd^{2} s^{2}+1)}}\). The correlation equation becomes \(XD_{a,b} Y^{\mathsf{T}}=-\zeta
I_{2}\), where
\begin{equation}
\label{eq:C.3}
D_{a,b}=\begin{pmatrix}
-\sigma_{0} & 0 & 0 \\
0 & \frac{\tilde{w}}{\sqrt{ab}} & \frac{\tilde{\beta}}{\sqrt{a}} \\
0 & \frac{\tilde{\gamma}}{\sqrt{b}} & 0
\end{pmatrix}.
\end{equation}
The scalar block represents the contribution from the response means, while the lower-right block \(B_{\ast}\)
couples the two-label fluctuations. Let \(\sigma_{+} \ge \sigma_{-}>0\) denote the singular values of
\(B_{\ast}\). If \(\sigma_{0}<\sigma_{-}\), the two largest singular values of \(D_{a,b}\) are \(\sigma_{+}\)
and \(\sigma_{-}\). The determinant and Frobenius norm of \(B_{\ast}\) give
\begin{equation*}
\sigma_{+} \sigma_{-}=\frac{\tilde{\beta}\tilde{\gamma}}{\sqrt{ab}},\quad (\sigma_{+}+\sigma_{-})^{2}=\frac{\tilde{w}^{2}+\left( \tilde{\gamma}\sqrt{a}+\tilde{\beta}\sqrt{b}\right)^{2}}{ab}.
\end{equation*}
Applying Eq.~\eqref{eq:C.2} therefore yields
\begin{equation}
\label{eq:C.4}
\zeta \le \sqrt{(1+2a)(1+2b)}\frac{\tilde{\beta}\tilde{\gamma}}{\sqrt{P_{+}}},\quad P_{+}=\tilde{w}^{2}+\left( \tilde{\gamma}\sqrt{a}+\tilde{\beta}\sqrt{b}\right)^{2}.
\end{equation}
For \(\frac{2}{5}\le s\le 1\), the weights \(a,b\) can be chosen so that the required singular-value ordering holds and the right-hand
side is at most \(Z(d)\).

\subsection{Uniform bound on the visibility}
\label{sec:C.3}

First consider \(0<s\le \frac{2}{5}\). Choose a unit vector perpendicular to \(\Delta v\), taking any unit
vector if \(\Delta v=0.\) Right-multiplying the correlation identity in Eq.~\eqref{eq:C.1} by this vector
eliminates the terms proportional to \((\Delta v)^{\mathsf{T}}\). The Cauchy--Schwarz inequality gives
\begin{equation*}
\zeta \le \sqrt{2}\sqrt{\left\Vert \overline{v}\right\Vert^{2}+\tilde{\beta}^{2} \left\Vert \Delta s\right\Vert^{2}}.
\end{equation*}
Since \(\tilde{\beta}\le ds\), the remaining norm constraint in Eq.~\eqref{eq:C.1} implies
\begin{equation*}
\zeta\le \frac{2\sqrt{2}}5\,d.
\end{equation*}
The function \(Z\) is concave on \([\frac{1}{2},1]\), since its radicand is positive and concave. The endpoint
values \(Z(\frac{1}{2})=\frac{1}{3}\) and \(Z(1)=\frac{1}{\sqrt{2}}\) therefore give \(Z(d)\ge\frac23d\), so \(\zeta<Z(d)\).

For \(\frac{2}{5}\le s\le 1\), set \(x=2d-1\) and choose \(a=\frac{x(1-x+xs)}{2d^{2}}\) and
\(b=\frac{x(1-x+\frac{x}{s})}{2d^{2}}\). At \(s=1\), the right-hand side of Eq.~\eqref{eq:C.4} equals
\(Z(d)\). Appendix~\ref{sec:C.4} proves \(\sigma_{0}<\sigma_{-}\) and
\begin{equation}
\label{eq:C.5}
R(d,s)=Z(d)^{2} P_{+}-(1+2a)(1+2b)(\tilde{\gamma}\tilde{\beta})^{2} \ge 0.
\end{equation}
Thus Eqs.~\eqref{eq:C.4} and \eqref{eq:C.5} give \(\zeta \le Z(d)\) for \(\frac{2}{5}\le s\le 1\). The
projection bound covers the remaining \(s\). Inverting \(Z\) with \(\delta=d^{2}\) gives
\begin{equation*}
\delta \ge \left( \sqrt{\frac{2}{1-\zeta^{2}}}-1\right)^{2},\frac{1}{3}<\zeta<\frac{1}{\sqrt{2}}.
\end{equation*}
The protocol in Appendix~\ref{sec:B.2} attains equality. For \(\zeta \ge \frac{1}{\sqrt{2}}\), the bound
\(Z(d)<\frac{1}{\sqrt{2}}\) for all \(d<1\) leaves pure QQ as the only feasible mode with at most two labels,
so \(q_{2} (\zeta)=1\).

\subsection{Auxiliary estimates}
\label{sec:C.4}

\subsubsection*{Nonnegative residual}

Weighted AM--GM applied separately to \(1-x+xs\) and \(1-x+\frac{x}{s}\) gives \(\sqrt{ab}\ge\frac{x}{2d^{2}}\). Substituting this lower bound into \(P_{+}\) and factoring the resulting rational expression yields
\begin{equation*}
R(d,s)\ge \frac{(1-d)x(1-s)^{2} [2(d^{2}+2d-1)(s-1)^{2}+sT_{0} (d)]}{2(d+1)^{2} (d+s)^{2} (1+ds)^{2}},
\end{equation*}
where \(T_{0}(d)=26d-10+(1-d)(2d-1)(7d^{2}+14d+1)\).
On \([\frac12,1]\), \(T_{0}(d)>0\) and all remaining factors are nonnegative.

\subsubsection*{Singular-value ordering}

The inequality \(\sigma_{0}<\sigma_{-}\) is equivalent to positive definiteness of \(B_{\ast}^{\mathsf{T}}
B_{\ast}-\sigma_{0}^{2} I_{2}\). Extend the matrix continuously to \(d=1\). On the connected domain \(\frac{1}{2}<d\le1\),
\(\frac{2}{5}\le s\le1\), it suffices to show that its determinant never vanishes and that the matrix is
positive definite at one point. Set \(\kappa=s+s^{-1}-2\in[0,1]\) and write
\begin{equation*}
F=2(ad^{2}+s^{2}),\quad G=2(bd^{2}s^{2}+1).
\end{equation*}
Both quantities are positive. Expanding the determinant gives
\begin{equation}
\label{eq:C.6}
ab\det \,(B_{\ast}^{\mathsf{T}} B_{\ast}-\sigma_{0}^{2} I_{2})=\frac{d^{4} s^{6} Q(d,\kappa)}{(d+s)^{2} (1+ds)^{2} F^{2} G^{2}},
\end{equation}
where
\begin{align}
Q&=1+2(1-d)[(x+2)(x+3)+x\kappa (f-4) \notag\\
 &\quad{}+2x\kappa^{2} (g-3)+2(1-x)x^{2} \kappa^{3} ((1-x)^{2}-3)], \notag\\
2f(x)&=2x^{3} (x+3)(x+6)-9x^{2}-43x+20, \notag\\
\label{eq:C.7}
4g(x)&=19-38x-10x^{2}+48x^{3} -3x^{4} [(1-x)^{2}+1].
\end{align}
For \(0\le x\le 1\), use \(18x^{4}-9x^{2} \ge-\frac{9}{8}\), \(x^{4} [(1-x)^{2}+1]\le 2x^{3}\), and \(2x^{2}
\le x+x^{3}\) in the expressions for \(f\) and \(g\). Then
\begin{equation*}
\begin{gathered}
\min \{2f(x),4g(x)\}\ge 36x^{3}-43x+\frac{75}{4} \\
=36(x-\frac{5}{8})^{2} (x+\frac{5}{4})+\frac{75-52x}{64}>0.
\end{gathered}
\end{equation*}
Since \(0\le\kappa\le1\), the bracket in \(Q\) is bounded below by \((1+x)[x+6(1-x)^2]>0\).
Hence \(Q\ge 1\) and the determinant in Eq.~\eqref{eq:C.6} is positive throughout the domain. At \(d=s=1\),
the matrix equals \(\frac{1}{18}I_{2}\). Its eigenvalues remain positive by continuity, so
\(\sigma_{-}^{2}>\sigma_{0}^{2}\) as required in Eq.~\eqref{eq:C.4}.

\section{Three-label lower bound}
\label{sec:D}

\subsection{Matching conditions and witness bounds}
\label{sec:D.1}

For \(\delta<1\), use the normalized branch weights \(\tilde{w}, \tilde{\beta},\tilde{\gamma}\) defined in
Appendix~\ref{sec:C.1}. Define the scalar mode asymmetry
\(\Delta=\tilde{\beta}-\tilde{\gamma}=\frac{\eta_{A}-\eta_{B}}{\tau}.\) The independence relation in
Eq.~\eqref{eq:4} becomes
\begin{equation}
\label{eq:D.1}
\delta \tilde{w}=(1-\delta)\tilde{\beta}\tilde{\gamma}.
\end{equation}
The matching conditions are
\begin{equation}
\label{eq:D.2}
\begin{gathered}
(\tilde{w}+\tilde{\beta})\overline{u}+\tilde{\gamma}\overline{r}=0,\quad (\tilde{w}+\tilde{\gamma})\overline{v}+\tilde{\beta}\overline{s}=0, \\
\tilde{w}\overline{uv^{\mathsf{T}}}+\tilde{\beta}\overline{us^{\mathsf{T}}}+\tilde{\gamma}\overline{rv^{\mathsf{T}}}=-\zeta I_{2}.
\end{gathered}
\end{equation}

For \(0\le \zeta \le \frac{2}{3}\), a pure-CC realization is
\begin{equation}
\label{eq:D.3}
\begin{gathered}
(p_{1},p_{2},p_{3})=\left( \frac{3}{8},\frac{3}{8},\frac{1}{4}\right), \\
u_{1}=\left(-\frac{1}{3},-1\right)^{\mathsf{T}},\quad u_{2}=\left( 1,\frac{1}{3}\right)^{\mathsf{T}},\quad u_{3}=(-1,1)^{\mathsf{T}},\quad v_{j}=-\frac{3\zeta}{2}u_{j}.
\end{gathered}
\end{equation}
The responses are admissible, their weighted means vanish, and
\(\overline{uu^{\mathsf{T}}}=\frac{2}{3}I_{2}\). Thus \(q_{3} (\zeta)=0\) throughout this interval.

For \(\zeta>\frac{2}{3}\), combining the two local-mean equations and the correlation-matrix equation in
Eq.~\eqref{eq:D.2} gives
\begin{align}
0&=\mathbb{E}[(1-\tilde{\gamma})h\cdot u+(1-\tilde{\beta})k\cdot v+\tilde{w}u^{\mathsf{T}} Wv \notag\\
 &\quad{}+\tilde{\gamma}r\cdot (h+Wv)+\tilde{\beta}s\cdot (k+W^{\mathsf{T}} u)]+\zeta \operatorname{Tr} \,W. \notag\\
\Phi (u,v)&=(1-\tilde{\gamma})h\cdot u+(1-\tilde{\beta})k\cdot v+\tilde{w}u^{\mathsf{T}} Wv \notag\\
\label{eq:D.4}
 &\quad{}+\zeta \operatorname{Tr} \,W-\tilde{\gamma}\Vert h+Wv\Vert-\tilde{\beta}\Vert k+W^{\mathsf{T}} u\Vert,\quad \mathbb{E}\Phi \le 0,
\end{align}
for arbitrary \(h,k\in \mathbb{R}^{2}\) and \(W\in \mathbb{R}^{2\times 2}\). The same witness \(\left(
h,k,W\right)\) is applied to every shared label of a protocol.

Assign each classical response vector to a closed quadrant containing it. With at most three shared labels, at
least one of the four quadrants is unused on each side. Flipping the same coordinates on both sides preserves
the target, so take \((+,+)\) as an unused quadrant on Bob's side. Use
\begin{equation*}
h=(-3,-3),\quad k=(-5,-5),\quad W=\begin{pmatrix}
10 & 2 \\
2 & 10
\end{pmatrix}.
\end{equation*}

At \(\zeta=\frac{2}{3}\), the exact preliminary verifier establishes \(\Phi>0\) throughout
\(\tilde{\gamma}\ge \frac{4}{5}\). It checks all response vertices
\begin{equation*}
u,v\in \{-1,0,1\}^{2},\quad v\ne (1,1),
\end{equation*}
at the three mode vertices
\begin{equation*}
(\tilde{\beta},\tilde{\gamma})=(0,1),\quad \left( 0,\frac{4}{5}\right),\quad \left( \frac{1}{5},\frac{4}{5}\right).
\end{equation*}
For this witness, \(\Phi\) is concave in \(u\) and \(v\) separately and affine in the normalized branch weights
\((\tilde{\beta},\tilde{\gamma})\). The vertex bounds therefore hold throughout the indicated region. Since \(\operatorname{Tr} \,W>0\), the bound remains
strictly positive for \(\zeta>\frac{2}{3}\), contradicting \(\mathbb{E}\Phi \le 0\). Exchanging the parties
gives
\begin{equation*}
\tilde{\beta},\tilde{\gamma}<\frac{4}{5}.
\end{equation*}
These bounds exclude \(\tilde{w}+\tilde{\beta}\tilde{\gamma}=0\). Solving Eq.~\eqref{eq:D.1} for \(\delta\) gives
\begin{equation}
\label{eq:D.5}
\delta=\frac{(1-\tilde{w})^{2}-\Delta^{2}}{(1+\tilde{w})^{2}-\Delta^{2}},\tilde{w}>0.
\end{equation}

\subsection{Geometric exclusion of small costs}
\label{sec:D.2}

Write
\begin{equation*}
S=[-1,1]^{2},\quad D_{r}=\{x:\Vert x\Vert_{1} \le r\}.
\end{equation*}

For \(\zeta>\frac{2}{3}\), every feasible protocol with at most three shared labels and \(\delta<1\)
satisfies \(\delta>\frac{1}{4096}\). This inequality sets the lower
cutoff for the global cover.

\subsubsection*{Triangle containment}

Let \(T\subset S\) be a triangle containing both \(D_{r}\) and \(tS\), with \(r>\frac{1}{2}\). Then \(r+t\le
1\).

Extend the vertices of \(T\) radially to the boundary of \(S\). If two vertices lie on the same
edge, replace them by its endpoints. These steps only enlarge the triangle. By symmetry, its vertices may now
be written as
\begin{equation*}
(-X,-1),\quad (1,y),\quad (-Y,1),
\end{equation*}
with \(Y\le X\le 1.\) Since \((-r,0)\in D_{r}\), the left edge gives \(X+Y\ge 2r\).

Suppose \(t>t_{0}=1-r\), and set \(h=X-r.\) Then \(0\le h\le t_{0}\) and \(Y\ge r-h.\) Both points
\((t_{0},\pm t_{0})\) lie strictly inside \(tS.\) Applying the oblique edge constraints to these two points
and to \((0,-r)\in D_{r}\) gives
\begin{equation*}
y\le \frac{t_{0}}{r+h}-r,y<\frac{r^{2}}{1+h}-t_{0},-y<\frac{r^{2}}{1-h}-t_{0},
\end{equation*}
where the last bound uses \(Y\ge r-h\). Adding the last inequality to each of the first two and clearing
positive denominators yields
\begin{equation*}
(h-t_{0})(h-t_{0}+r^{2})>0, t_{0}-r^{2}<t_{0} h^{2}.
\end{equation*}
Since \(h\le t_{0}\), the first inequality gives \(h<t_{0}-r^{2}\). The second gives \(t_{0}-r^{2}<h\), a contradiction. Hence \(r+t\le 1\).

\subsubsection*{Clipped-square consequence}

Let \(O_{R}=\{x:\Vert x\Vert_{\infty} \le 1,\Vert x\Vert_{1} \le R\},\) where \(\sqrt{2}\le R\le 2.\) Any
triangle \(T\) satisfying \(D_{r} \subseteq T\subseteq O_{R}\) obeys
\begin{equation*}
r\le \frac{R}{1+R}.
\end{equation*}
For \(r>\frac{1}{2}\), polarity gives \(D_{r} \cup \frac{r}{R}S\subseteq rT^{\circ} \subseteq S\). Applying
the preceding triangle bound to \(rT^{\circ}\) yields \(r+\frac{r}{R}\le 1\). For \(r\le \frac{1}{2}\), the
result follows from \(R\ge 1.\)

\subsubsection*{Application to the effective response triangle}

The effective responses \(\xi_{j},\chi_{j}\) in Eq.~\eqref{eq:B.2} realize \((1-\delta)P_{\zeta}+\delta
P_{\mathrm{RS}}\), where \(P_{\mathrm{RS}}\) is generated by the local quantum responses \(r_{j},s_{j}\) with label
probabilities \(p_{j}\). Set \(d=\sqrt{\delta}\) and \(k_{0}=2-\sqrt{2}\). Exchange the parties if necessary
so that \(1-\eta_{A} \ge d\). Since \(\Vert \xi_{j} \Vert_{\infty} \le 1\) and
\begin{equation*}
\Vert \xi_{j}\Vert_{1}\le 2-k_{0}d,
\end{equation*}
Alice's response triangle \(T=\operatorname{conv} \{\xi_{j} \}\) lies in \(O_{2-k_{0} d}\).

For any \(h\ne 0\), choose Bob's input \(y\) and output \(b_{0}\) so that \(-b_{0} h_{y}=\Vert
h\Vert_{\infty}\). For the mixture \((1-\delta)P_{\zeta}+\delta P_{\mathrm{RS}}\), Alice's conditional mean vector \(m\) lies in \(T\),
with label weights proportional to \(p_{j}(1+b_{0}\chi_{jy})\). Since Bob's marginal under \(P_{\zeta}\)
is uniform and \(P_{\mathrm{RS}}(b_{0}\mid y)\le1\), the conditional weight \(\omega\) of \(P_{\zeta}\) satisfies
\begin{equation*}
\omega\ge \frac{1-\delta}{1+\delta}.
\end{equation*}
Under \(P_{\zeta}\), Alice's conditional mean vector given Bob's input \(y\) and outcome \(b_{0}\) is
\(-b_{0}\zeta e_{y}\), where \(e_{y}\) is the \(y\)th coordinate vector. The
local quantum responses lie in the unit disk, so \(\Vert h\Vert_{2} \le \sqrt{2}\Vert h\Vert_{\infty}\) gives
\begin{equation*}
h\cdot m\ge \rho\Vert h\Vert_{\infty},\quad \rho=\frac{(1-\delta)\zeta-2\sqrt{2}\delta}{1+\delta}.
\end{equation*}
For \(\rho>0\), the support function of \(D_{\rho}\) is \(\rho \Vert h\Vert_{\infty}\). Thus \(D_{\rho}
\subseteq T\).

Suppose \(d\le \frac{1}{64}\). Since \(\zeta>\frac{2}{3}\), the definition of \(\rho\) gives
\(\rho>\frac{2}{3}-cd^{2}\) with \(c=\frac{4}{3}+2\sqrt{2}\). This lower bound is positive, so the clipped-square bound \(\rho\le\frac23-\frac{k_0d}{9}\) yields
\begin{equation}
\label{eq:D.6}
cd^{2}>\frac{k_{0}d}{9}.
\end{equation}
This excludes \(d=0\) and otherwise requires \(d>\frac{1}{64}\).
Hence \(\delta>\frac{1}{4096}\).

\subsection{Exact global localization}
\label{sec:D.3}

For \(\zeta>\frac{2}{3}\) and \(\delta<1\), only candidates with \(\delta \le q_{2} (\zeta)\) can match or
improve on the attainable two-label benchmark. Put \(\delta_{\ast}=q_{2} (\frac{2}{3})\). Inverting
Eq.~\eqref{eq:24} gives the necessary visibility floor
\begin{equation*}
\zeta \ge g(\delta),\quad g(\delta)=\begin{cases}
\frac{2}{3}, & \mspace{-8mu}0\le \delta \le \delta_{\ast}, \\[-1pt]
\sqrt{1-\frac{2}{\left( 1+\sqrt{\delta}\right)^{2}}}, & \mspace{-8mu}\delta_{\ast}<\delta<1.
\end{cases}
\end{equation*}
The cost \(\delta=\frac{\tilde{\beta}\tilde{\gamma}}{(1-\tilde{\beta})(1-\tilde{\gamma})}\) is nondecreasing
in each coordinate on \(\tilde{\beta},\tilde{\gamma}\ge 0\), \(\tilde{\beta}+\tilde{\gamma}<1\). A rational
lower bound on \(g(\delta)\) at the lower corner of a mode cell therefore applies throughout the cell.

\begin{appendixlemma}[Global localization]
\label{lem:D.1}
\normalfont
Every feasible protocol using at most three labels with \(\zeta>\frac{2}{3},\delta<1,\) and \(\delta \le q_{2}
(\zeta)\) satisfies
\begin{equation}
\label{eq:D.7}
\delta>\frac{4}{5},\quad|\Delta|\le \frac{1}{20},\quad 0<\tilde{w}\le A<\frac{7}{125},\quad A=\frac{1-\sqrt{\delta}}{1+\sqrt{\delta}}.
\end{equation}
\end{appendixlemma}
Here \(A\) is the normalized CC weight of the symmetric mode pair at the same cost. Up to target-preserving
symmetries, five families of closed sign cells remain. With
\begin{equation*}
e=\frac{(1,1)}{\sqrt{2}},\quad f=\frac{(1,-1)}{\sqrt{2}},
\end{equation*}
the corresponding closed cells have principal corners \(\pm \sqrt{2}(e,f)\) and one of the third corners
\begin{equation*}
\sqrt{2}(e,-f),\quad \sqrt{2}(e,e),\quad \sqrt{2}(e,-e),\quad \sqrt{2}(f,e),\quad \sqrt{2}(f,-e).
\end{equation*}

\subsubsection*{Grid relaxation and coverage}

A grid row is a pair \(u,v\in \{-1,0,1\}^{2}\). It is complete if both vectors are square corners and
incomplete otherwise. For a label in a closed sign cell, write \(q=(u_{0},u_{1},v_{0},v_{1})\) and let
\(\sigma_{k}\) be the sign assigned to coordinate \(k\). Interpolate the coordinates independently, assigning
weight \(|q_{k}|\) to \(\sigma_{k}\) and \(1-|q_{k}|\) to zero. These weights reproduce every \(q_{k}\) and
\(u_{x} v_{y}\), including on cell boundaries. With the quantum responses fixed, the interpolation preserves
each label's contribution to Eq.~\eqref{eq:D.2}. For a fixed witness, separate concavity gives \(\Phi (q)\ge
\mathbb{E}_{\mathrm{grid}}[\Phi(Q)]\), where \(Q\) is the grid row sampled by interpolation at fixed \(q\). A lower bound on all allowed rows therefore
holds throughout the cell.

Number the quadrants \((-,-),\allowbreak(-,+),\allowbreak(+,-),\allowbreak(+,+)\) from 0 to 3, respectively,
and encode the joint cell \((\alpha,\beta)\) by \(4\alpha+\beta\). Under simultaneous signed coordinate
permutations and exchange of the parties, the 560 three-cell sets form 49 orbits. The mode cover excludes 43
and retains five for Appendix~\ref{sec:D.4}. Every smaller support extends to a covered three-cell set outside
the remaining orbit, represented by \(\{3,6,9\}\).

For the representative \(\{3,6,9\}\), each of the three occupied cells contains one label. When \(\tilde{w}\le
\frac{1}{4}\), the witness
\begin{equation*}
h=(-6,-6),\quad k=(6,6),\quad W=\begin{pmatrix}
42 & 7 \\
7 & 42
\end{pmatrix}
\end{equation*}
gives \(\Phi>0\) at every response vertex for \(\zeta=\frac{2}{3}\), excluding this range of
\(\tilde{w}\). For \(\tilde{w}>\frac{1}{4}\), subdivision in the twelve original response coordinates and the two
mode coordinates before grid expansion excludes this orbit as well. For all other orbits, the sign cells are kept whole and only the two
mode coordinates are subdivided.

For the support \(\left\{ 0,3,12\right\}\), the entire mode domain is excluded by the witness
\begin{equation*}
h=k=0,\quad W=\begin{pmatrix}
16 &-7 \\
-7 & 16
\end{pmatrix}.
\end{equation*}
Each grid vector belongs to \(\left\{ 0,\pm \left( 1,0\right),\pm \left( 0,1\right),\pm \left(
1,1\right)\right\}\). On these vectors, \(\left| u^{\mathsf{T}} Wv\right| \le 18\) and \(\left\Vert Wu\right\Vert,\left\Vert Wv\right\Vert <18\). Since \(\operatorname{Tr} \,W=32\) and
\(\tilde{w}+\tilde{\beta}+\tilde{\gamma}=1\), Eq.~\eqref{eq:D.4} gives \(\Phi>0\) on every grid row and hence, by interpolation, throughout each cell. This contradicts \(\mathbb{E}\Phi\le 0\).

The verifier checks that the closed subregions cover each parent, including shared boundaries. Each terminal
region is retained for Appendix~\ref{sec:D.4} or excluded by a domain restriction, symmetry, an analytic
bound, or a single witness valid throughout the region. When \(\operatorname{Tr} \,W\ge 0\), a positive lower
bound on \(\Phi\) at the visibility floor suffices for exclusion. At a floor of \(\frac{2}{3}\), a zero lower
bound also suffices if \(\operatorname{Tr} \,W>0\), since every candidate satisfies \(\zeta>\frac{2}{3}\). The
checks use exact rational arithmetic and integer-square bounds. Reference \cite{Verification31} gives the full
partition and witness parameters.

\subsection{Averaged local estimate}
\label{sec:D.4}

On the retained region, define
\begin{equation*}
Z_{\tilde{w}}=\sqrt{\frac{1}{2}-\tilde{w}-\frac{\tilde{w}^{2}}{2}},\quad n=\frac{\tilde{w}}{Z_{\tilde{w}}},\quad F(x)=\sqrt{\left( 1+x\right)^{2}+n^{2}},\quad|x|\le \frac{1}{20}.
\end{equation*}

\begin{appendixlemma}[Averaged local bound]
\label{lem:D.2}
\normalfont
Every feasible candidate retained by Lemma~\ref{lem:D.1} satisfies
\begin{equation}
\label{eq:D.8}
2\zeta \le \frac{1-\tilde{w}}{\sqrt{2}}\left[ F\left( \Delta\right)+F\left(-\Delta\right)\right]-2\tilde{w}n+\frac{\left( 1-2w_{3}\right) \Delta}{\sqrt{2}}\left[ F\left(-\Delta\right)-F\left( \Delta\right)\right].
\end{equation}
\end{appendixlemma}

In the two exchanged-axis families, \(w_{3} \in [0,\frac{7}{19}]\) denotes the weight of the exceptional grid
corner. In the other three families, dropping the positive third-row gap gives Eq.~\eqref{eq:D.8} with
\(w_{3}=0\).

Set \(R=\sqrt{2}F(\Delta)\), \(S=\sqrt{2}F(-\Delta)\), and
\(L=2\zeta+2\tilde{w}n-\tilde{\gamma}R-\tilde{\beta}S\). Table~\ref{tab:2} lists one witness per family in the
orthonormal \((e,f)\) basis of Appendix~\ref{sec:D.3}. For the first three rows, define
\begin{equation*}
H_{b}=\tilde{\beta}S-2(1-\tilde{\beta})n,\quad H_{c}=\tilde{\gamma}R-2(1-\tilde{\gamma})n,
\end{equation*}

\begin{equation*}
H=H_{b}+H_{c},\quad \ell=\frac{\kappa H_{b}}{H},\quad t=\frac{\kappa H_{c}}{H}.
\end{equation*}
Both \(H_{b}\) and \(H_{c}\) are positive on the closed rectangle \(0\le \tilde{w}\le \frac{7}{125}\),
\(|\Delta|\le \frac{1}{20}\). Reference \cite{Verification31} verifies the row bounds below directly on this
rectangle.

\begin{table}[!htb]
  \centering
  \caption{Explicit witnesses for the five retained sign-cell families. The actual third corner is \(\sqrt{2}\) times the
listed pair.}
  \label{tab:2}
  \small
  \setlength{\tabcolsep}{6pt}
  \renewcommand{\arraystretch}{1.55}
  \begin{tabular}{|Gc|Gc|Gc|Gc|}
    \hline
    \textbf{Third pair} & \(\frac{h}{\sqrt{2}}\) & \(\frac{k}{\sqrt{2}}\) & \(W\) \\
    \hline
    \shortstack{\((e,-f)\) \\ \(\kappa=\frac{1}{16}\)} & \(\ell \begin{pmatrix}
n \\
1+\Delta
\end{pmatrix}\) & \(-t\begin{pmatrix}
1-\Delta \\
n
\end{pmatrix}\) & \(\begin{pmatrix}
1-\Delta & n \\
0 & 1+\Delta
\end{pmatrix}\) \\
    \hline
    \shortstack{\((e,e)\) \\ \(\kappa=\frac{9}{20}-3\tilde{w}\)} & \(-\ell \begin{pmatrix}
n \\
1+\Delta
\end{pmatrix}\) & \(t\begin{pmatrix}
1-\Delta \\
n
\end{pmatrix}\) & \(\begin{pmatrix}
1-\Delta & n \\
\ell (1+\Delta) & 1+\Delta
\end{pmatrix}\) \\
    \hline
    \shortstack{\((e,-e)\) \\ \(\kappa=\frac{1}{4}\)} & \(\ell \begin{pmatrix}
n \\
1+\Delta
\end{pmatrix}\) & \(-t\begin{pmatrix}
1-\Delta \\
n
\end{pmatrix}\) & \(\begin{pmatrix}
1-\Delta & n \\
\ell (1+\Delta) & 1+\Delta
\end{pmatrix}\) \\
    \hline
    \((f,e)\) & \(0\) & \(0\) & \(\begin{pmatrix}
1-\Delta & n \\
n & 1+\Delta
\end{pmatrix}\) \\
    \hline
    \((f,-e)\) & \(0\) & \(0\) & \(\begin{pmatrix}
1-\Delta & n \\
-n & 1+\Delta
\end{pmatrix}\) \\
    \hline
  \end{tabular}
\end{table}

The two principal complete rows have \(\Phi=L\). Every incomplete row has \(\Phi>L\), as does the third
complete row in each of the first three families. In the two exchanged-axis families, the third complete
row has \(\Phi=L+\Delta(S-R)\). For the \(\left( f,e\right)\) witness in Table~\ref{tab:2},
substitution of \(u=\sqrt{2}e,v=\sqrt{2}f\) gives \(u^{\mathsf{T}} Wv=2n\), \(\left\Vert Wv\right\Vert=R\),
and \(\left\Vert W^{\mathsf{T}} u\right\Vert=S\).

Since both auxiliary witnesses in Table~\ref{tab:3} have positive trace, their strict row bounds hold
for all \(\zeta\ge\frac{2}{3}\). Averaging these bounds and using \(\mathbb{E}\Phi\le0\) gives
\(w_{3}<\frac{7}{19}\) in both exchanged-axis families.

For the witness assigned to each family in Table~\ref{tab:2}, the row bounds and \(\mathbb{E}\Phi\le0\) give
\begin{equation*}
L+w_{3}\Delta(S-R)\le0.
\end{equation*}
Substituting \(L,R,S\) and using \(\Delta=\tilde{\beta}-\tilde{\gamma}\) gives Eq.~\eqref{eq:D.8}. For \(n<\frac{1}{11}\), the derivatives satisfy
\begin{equation*}
F'(x)>\frac{99}{100},0\le F''(x)<\frac{1}{100},|x|\le \frac{1}{20}.
\end{equation*}
Integrating these bounds yields
\begin{equation*}
F(\Delta)+F(-\Delta)\le 2F(0)+\frac{\Delta^{2}}{100},\quad \Delta [F(-\Delta)-F(\Delta)]\le-\frac{99}{50}\Delta^{2}.
\end{equation*}
Using \(\frac{(1-\tilde{w})F(0)}{\sqrt{2}}-\tilde{w}n=Z_{\tilde{w}}\) and \(w_{3} \le \frac{7}{19}\) in
Eq.~\eqref{eq:D.8} yields
\begin{equation}
\label{eq:D.9}
\zeta \le Z_{\tilde{w}}-\frac{7}{40}\Delta^{2}.
\end{equation}

\begin{table}[H]
  \centering
  \caption{Auxiliary witnesses in the original \(X/Z\) coordinates and strict lower bounds on \(\Phi\). Third pairs are
normalized as in Table~\ref{tab:2}. The last two columns refer to ordinary and exceptional grid rows.}
  \label{tab:3}
  \small
  \setlength{\tabcolsep}{6pt}
  \renewcommand{\arraystretch}{1.55}
  \begin{tabular}{|Gc|Gc|Gc|Gc|Gc|Gc|}
    \hline
    \textbf{Third pair} & \(h\) & \(k\) & \(W\) & \textbf{Other rows} & \textbf{Third row} \\
    \hline
    \((f,e)\) & \((6,-1)^{\mathsf{T}}\) & \((6,1)^{\mathsf{T}}\) & \(\begin{pmatrix}
9 &-1 \\
1 & 10
\end{pmatrix}\) & \(-\frac{7}{5}\) & \(\frac{12}{5}\) \\
    \hline
    \((f,-e)\) & \((2,-2)^{\mathsf{T}}\) & \((-2,-2)^{\mathsf{T}}\) & \(\begin{pmatrix}
4 & 0 \\
0 & 4
\end{pmatrix}\) & \(-\frac{3}{2}\) & \(3\) \\
    \hline
  \end{tabular}
\end{table}

\subsection{Optimal cost and endpoints}
\label{sec:D.5}

Write \(Z_{A}=Z(\sqrt{\delta})\) as in Eq.~\eqref{eq:B.6}. Rationalizing \(Z_{\tilde{w}}-Z_A\) and using Eq.~\eqref{eq:D.5} gives the upper bound \(\frac{A\Delta^2}{2(1-A)Z_A}\). Since Eq.~\eqref{eq:D.7} gives \(A<\frac{7}{125}\) and \(Z_A>\frac35\),
\begin{equation*}
Z_{\tilde w}-Z_A\le\frac{\Delta^2}{20}.
\end{equation*}
Combining this bound with Eq.~\eqref{eq:D.9} yields
\begin{equation}
\label{eq:D.10}
\zeta \le Z(\sqrt{\delta})-\frac{\Delta^{2}}{8}.
\end{equation}

For \(\frac{2}{3}<\zeta<\frac{1}{\sqrt{2}}\), strict monotonicity gives \(Z_A\le\zeta\) whenever \(\delta \le q_{2} (\zeta)\). Eq.~\eqref{eq:D.10} then forces \(\delta=q_{2}
(\zeta)\) and \(\Delta=0\). The construction in Appendix~\ref{sec:B.2} attains this cost, and every optimum
has \(\eta_{A}=\eta_{B}\).

For \(\zeta \ge \frac{1}{\sqrt{2}}\), any candidate with \(\delta<1\) would have \(Z_{A}<\frac{1}{\sqrt{2}},\)
contradicting Eq.~\eqref{eq:D.10}. Pure QQ attains unit cost. Together with the zero-cost construction in
Eq.~\eqref{eq:D.3} for \(\zeta \le \frac{2}{3}\), this proves Eq.~\eqref{eq:25a}.

\subsection{Equality and effective label number}
\label{sec:D.6}

For \(\frac{2}{3}<\zeta<\frac{1}{\sqrt{2}}\), Appendix~\ref{sec:D.5} gives \(\Delta=0\) and
\(\zeta=Z_{\tilde{w}}\) at every optimum, so \(L=0\). The row bounds in Appendix~\ref{sec:D.4} are then
nonnegative and are strictly positive on every incomplete row and on the third complete row in the first three
families. Since \(\mathbb{E}\Phi \le 0\), these rows carry zero weight. Both classical responses of each
positive-weight label therefore lie at corners of the square, since the interpolation in
Appendix~\ref{sec:D.3} would otherwise give positive weight to an incomplete row.

Let \(p_{\pm}\) be the weights at the principal corners \(\pm \sqrt{2}(e,f)\). In either exchanged-axis
family, write \(w_{3}\) for the weight at \(\sqrt{2}(f,\varepsilon e)\), where \(\varepsilon=\pm 1\). For the
corresponding witness in Table~\ref{tab:2}, equality gives \(r=-\frac{Wv}{\left\Vert Wv\right\Vert}\). Substituting \(\frac{1-\tilde{w}}{\sqrt{2}\zeta}\) for \(\sqrt{1+n^{2}}\) in the \(e\) and \(f\) components of Alice's zero-mean condition gives
\begin{equation*}
p_{+}-p_{-}=\varepsilon \zeta w_{3},\quad w_{3}=\zeta \left( p_{+}-p_{-}\right).
\end{equation*}
Since \(\zeta^{2}<1\), these identities imply \(w_{3}=0\). In all five families, only the principal pair
remains, and Alice's vanishing local means and normalization give \(p_{+}=p_{-}=\frac{1}{2}\). Equality in the
disk bounds fixes the quantum responses at both corners. After identical tuples are merged, every optimum is
therefore the two-label protocol of Appendix~\ref{sec:B.2} up to target-preserving symmetries.

At \(\zeta=\frac{2}{3}\), Eq.~\eqref{eq:D.3} gives a pure-CC optimum with three labels. For \(\zeta \ge
\frac{1}{\sqrt{2}}\), every feasible protocol with at most three labels has \(\delta=1\), leaving its unused non-QQ responses arbitrary.

\section{Four-label modes and set saturation}
\label{sec:E}

\subsection{Complete four-label feasible mode set}
\label{sec:E.1}

The square and unit-disk response domains give, for each label, \(-u_{j} \cdot v_{j} \le 2\), \(-u_{j} \cdot
s_{j} \le \sqrt{2}\), and \(-r_{j} \cdot v_{j} \le \sqrt{2}\). Taking the negative trace of the non-QQ
correlation equation gives
\begin{equation}
\label{eq:E.1}
\tau \zeta \le w+\frac{\beta+\gamma}{\sqrt{2}}.
\end{equation}
This bound is necessary for every label budget.

To establish sufficiency with four labels, suppose \(\tau>0\) and set \(\theta=\frac{\sqrt{2}\zeta \tau}{\sqrt{2}w+\beta+\gamma}\).
The denominator is positive, and Eq.~\eqref{eq:E.1} ensures \(0\le \theta \le 1\). Choose four equally
weighted labels \(\varepsilon \in \{-1,+1\}^{2}\), with responses
\begin{equation}
\label{eq:E.2}
u_{\varepsilon}=\varepsilon,\quad v_{\varepsilon}=-\theta \varepsilon,\quad r_{\varepsilon}=\frac{\varepsilon}{\sqrt{2}},\quad s_{\varepsilon}=-\frac{\theta \varepsilon}{\sqrt{2}}.
\end{equation}
All responses satisfy the classical and quantum constraints. Since \(\overline{\varepsilon}=0\) and
\(\overline{\varepsilon \varepsilon^{\mathsf{T}}}=I_{2}\), all local means vanish, while the non-QQ
correlation contribution is \(-\tau \zeta I_{2}\). At \(\tau=0\), the protocol is pure QQ. Thus every mode pair satisfying Eq.~\eqref{eq:E.1} can be realized with at most four labels,
so \(\mathcal{F}_{4}=\mathcal{F}_{\infty}\).

It remains to express Eq.~\eqref{eq:E.1} in terms of \((\eta_{A},\eta_{B})\). For \(\zeta \le
\frac{1}{\sqrt{2}}\), it holds throughout \([0,1]^{2}\). For \(\zeta>\frac{1}{\sqrt{2}}\), substituting the
branch weights and using \(r_{\zeta}\) from Eq.~\eqref{eq:27} gives \((1+r_{\zeta})\eta_{A} \eta_{B} \ge
r_{\zeta} (\eta_{A}+\eta_{B})\). Apart from pure QQ, this requires \(\eta_{A},\eta_{B} \ge r_{\zeta}\). Thus
\((1+r_{\zeta})\eta_{A}-r_{\zeta}>0\), and solving for \(\eta_{B}\) gives Eq.~\eqref{eq:28}.

\subsection{Three-label pure-CQ geometry}
\label{sec:E.2}

For \(\zeta>0,\) a three-label pure-CQ realization exists if and only if there is a triangle \(T\) satisfying
\(\zeta B_{2} \subseteq T\subseteq [-1,1]^{2}\), where \(B_{2}\) is the unit disk. To establish necessity, let
\(T\) be the convex hull of the classical responses \(u_{j}\). The matching conditions are
\(\overline{u}=\overline{s}=0,\overline{us^{\mathsf{T}}}=-\zeta I_{2}\). For every unit vector \(n\), the
coefficients \(p_{j} (1+s_{j} \cdot n)\) are nonnegative and sum to one. Moreover,
\begin{equation*}
\sum_{j} p_{j} (1+s_{j} \cdot n)u_{j}=-\zeta n.
\end{equation*}
Thus \(T\) contains the circle of radius \(\zeta\) and, by convexity, the disk \(\zeta B_{2}.\)

Conversely, suppose such a triangle exists, with vertices \(u_{j}\) and affine barycentric coordinates
\(\lambda_{j} (x)=p_{j}+g_{j} \cdot x\). Since the origin lies in its interior, \(p_{j}=\lambda_{j} (0)>0\).
Nonnegativity of \(\lambda_{j}\) on \(\zeta B_{2}\) gives \(p_{j} \ge \zeta \Vert g_{j} \Vert.\) Hence
\(s_{j}=-\frac{\zeta g_{j}}{p_{j}}\) satisfies \(\Vert s_{j} \Vert \le 1.\) The barycentric identities then
give \(\overline{u}=\overline{s}=0,\overline{us^{\mathsf{T}}}=-\zeta I_{2}\), establishing sufficiency.

The parent-measurement correspondence in Proposition~\ref{prop:1} follows from the same moment conditions.
Given a pure-CQ realization, define
\begin{equation*}
G_{j}=p_{j} (I-s_{j0} X-s_{j1} Z),\quad t(a|x,j)=\frac{1+au_{jx}}{2}.
\end{equation*}
The bounds \(\Vert s_{j} \Vert \le 1\) and \(|u_{jx}|\le 1\) ensure positive effects and valid response
probabilities. The zero means imply \(\sum_{j} G_{j}=I\), and the correlation conditions give
\begin{equation*}
\sum_{j} t(a|x,j)G_{j}=\frac{I+a\zeta A_{x}}{2},\quad A_{0}=X, A_{1}=Z.
\end{equation*}

Conversely, write each nonzero parent effect as \(G_{j}=p_{j} (I+\mathbf{b}_{j} \cdot \boldsymbol{\sigma})\).
Projecting \(\mathbf{b}_{j}\) onto the \(X/Z\) plane preserves positivity and the two child measurements.
Setting \(s_{j}=-(b_{jX},b_{jZ})\) and \(u_{jx}=2t(+1|x,j)-1\) then recovers the pure-CQ moment conditions.

For the construction of Ref.~\cite{Skrzypczyk2020}, write \(\rho=\zeta_{\ast}\). This threshold lies between
\(\frac{1}{2}\) and \(\frac{1}{\sqrt{2}}\) and satisfies
\begin{equation}
\label{eq:E.3}
\rho^{6}+4\rho^{4}+5\rho^{2}=2,\quad \rho (\rho^{2}+3)=\sqrt{2}(1+\rho^{2}).
\end{equation}
Set \(a=\sqrt{2}\rho-1\). The triangle with vertices \((-1,-1),\allowbreak(1,a),\allowbreak(a,1)\) contains
the origin. The side joining the last two vertices has distance \(\rho\) from it. The other two have distance
\(\frac{2-\sqrt{2}\rho}{\sqrt{4+2\rho^{2}}}=\rho\) by Eq.~\eqref{eq:E.3}. This triangle realizes pure CQ for
every \(0<\zeta \le \rho\).

\subsection{Sharp upper bound on the disk radius}
\label{sec:E.3}

Let \(\frac{1}{2}<r<1\) and suppose a triangle in the square contains the centered disk of radius \(r\). By
the enlargement and symmetry argument in Appendix~\ref{sec:D.2}, it suffices to consider a triangle with
vertices \((-X,-1),\allowbreak(1,y),\allowbreak(-Y,1)\), where \(X\ge Y\). Put \(t=\frac{X-Y}{2}\) and
\(c=r\sqrt{1+t^{2}}\). The left side has distance at least \(r\), so \(X+Y\ge 2c\) and \(0<Y\le X\le 1\).
Move its endpoints to \(A_{\varepsilon}=(\varepsilon t-c,\varepsilon)\), \(\varepsilon=\pm 1\). This shifts
the side right to tangency and widens the other two half-planes for \(x\le 1\). The disk still fits and
\(c+t\le 1\).

In coordinates \(z=x-ty+c\), the second tangent through \(A_{\varepsilon}\) has equation
\(y=\varepsilon+k_{\varepsilon} z\). The slope is determined by requiring the line to lie at Euclidean distance \(r\) from the disk center
\((x,y)=(0,0)\). Solving the two tangent equations then gives their intersection abscissa \(x_{\ast}\).
\begin{equation}
\label{eq:E.4}
k_{\varepsilon}=\frac{r^{2}-1}{2(\varepsilon c-r^{2} t)},\quad x_{\ast}=\frac{r^{2} (1+r^{2})}{(1-r^{2})c}.
\end{equation}

Toward the right, the tangent from \(A_{+}\) descends and that from \(A_{-}\) rises. At \(x=1\) the apex lies
on or above the first and on or below the second, so \(x_{\ast} \le 1\). Eq.~\eqref{eq:E.4} gives the first
inequality in Eq.~\eqref{eq:E.5}, and the second follows from \(c+t\le 1\) and the definition of \(c\).
Comparing the outer terms and squaring yields the polynomial bound.
\begin{equation}
\label{eq:E.5}
r^{2} (1+r^{2})\le (1-r^{2})c\le r(\sqrt{2-r^{2}}-r),\quad r^{6}+4r^{4}+5r^{2} \le 2.
\end{equation}
By monotonicity, Eqs.~\eqref{eq:E.3} and \eqref{eq:E.5} rule out centered disks with radii in
\((\zeta_{\ast},1)\) and hence the unit disk, which contains them all. Exchanging the parties gives the same
threshold for pure QC.

\subsection{A three-label construction for every mode below the threshold}
\label{sec:E.4}

Use the three-label realization from Appendix~\ref{sec:E.2} at \(\rho=\zeta_{\ast}\). Write \(U_{j}=u_{j}\)
and \(R_{j}=-s_{j}\), so \(\mathbb{E}U=\mathbb{E}R=0\) and \(\mathbb{E}[UR^{\mathsf{T}}]=\rho I_{2}\).
Swapping the two coordinates permutes the vertices of the triangle and their barycentric functions. Thus the
joint law of \((U,R)\) is invariant under this swap.

For \(\tau>0\), put \(t=\frac{w}{\tau}\) and \(X_{j}=\frac{U_{j}+tR_{j}}{1+t}\). Before Bob's transformation,
both parties use \(X_j\) and \(R_j\) as their classical and quantum responses, respectively. The vectors \(X_j\) lie in the square. Since
\((1+t)\mathbb{E}[XR^{\mathsf{T}}]=\rho I_{2}+t\mathbb{E}[RR^{\mathsf{T}}]\) is symmetric, the CQ and QC branch
correlation matrices coincide before Bob's transformation. Let \(M\) denote the normalized non-QQ
correlation matrix at this stage. Expansion gives
\begin{equation}
\label{eq:E.6}
M-\rho I_{2}=\frac{t}{(1+t)^{2}}\mathbb{E}[(U-R)(U-R)^{\mathsf T}].
\end{equation}

Thus \(M\succeq\rho I_2\). For \(0<\zeta \le \rho\), let \(L=\zeta M^{-1}\). It commutes with the coordinate swap. Let \(a\) and \(b\) be the diagonal and off-diagonal entries of \(L\),
respectively. Its eigenvalues \(a+b\) and \(a-b\) lie in \([0,1]\), so \(\lvert a\rvert+\lvert b\rvert\le1\). Thus \(L\) preserves both the square and the unit disk. With Bob's
responses \(v_{j}=-LX_{j}\) and \(s_{j}=-LR_{j}\), the local means vanish and the normalized non-QQ correlation matrix is
\(-ML^{\mathsf{T}}=-\zeta I_{2}\), for any split of the two mixed branches. The endpoint \(\tau=0\) is pure
QQ.

\subsection{Necessity of the fourth label above the threshold}
\label{sec:E.5}

For \(\zeta_{\ast}<\zeta \le \frac{1}{\sqrt{2}}\), pure CQ is feasible with four labels by
Appendix~\ref{sec:E.1} and impossible with three by Appendices~\ref{sec:E.2} and \ref{sec:E.3}. For
\(\frac{1}{\sqrt{2}}<\zeta \le 1\), consider \(\eta_{A}=1\), \(\eta_{B}=r_{\zeta}\), with \(r_{\zeta}\) from
Eq.~\eqref{eq:27}. This boundary mode is feasible and saturates the bound in Eq.~\eqref{eq:E.1}. Since
\(w>0\), equality in the per-label trace bounds forces \(u_{j} \in \{-1,+1\}^{2}\) and \(v_{j}=-u_{j}\)
whenever \(p_{j}>0\). If \(\beta>0\), it also requires \(s_{j}=-\frac{u_{j}}{\sqrt{2}}\). The non-QQ
contribution is therefore \(-\left( w+\frac{\beta}{\sqrt{2}}\right) \overline{uu^{\mathsf{T}}}\). The zero
local means and vanishing cross-input correlations imply
\begin{equation}
\label{eq:E.7}
\overline{u_{0}}=\overline{u_{1}}=\overline{u_{0} u_{1}}=0.
\end{equation}

As in Appendix~\ref{sec:A.6}, Eq.~\eqref{eq:E.7} fixes each corner probability to \(\frac{1}{4}\),
so at least four shared labels are necessary. At \(\zeta=1\), the mixed-branch weight vanishes and the same
argument applies to the pure-CC mode.

For \(0<\zeta \le \zeta_{\ast}\), the pure-CC mode is feasible by Appendix~\ref{sec:D.1} but cannot be
realized with two labels by Appendix~\ref{sec:B.1}. Together with the construction in Appendix~\ref{sec:E.4},
this shows that three labels are necessary and sufficient in this interval. At \(\zeta=0\), a single
zero-response label realizes every mode. Combined with the three-label bounds in Appendix~\ref{sec:D}, these
results establish both saturation capacities in Fig.~\ref{fig:3}(a), including the threshold endpoints.

\section{Local marginals and source-cost separation}
\label{sec:F}

\subsection{Constructive proof of the two-label rank criterion}
\label{sec:F.1}

For any two-label pure-CC realization with probabilities \(p_{1},p_{2}\), the covariance identity gives
\begin{equation}
\label{eq:F.1}
\Gamma=p_{1} p_{2} (u_{1}-u_{2})(v_{1}-v_{2})^{\mathsf{T}}.
\end{equation}
Thus \(\operatorname{rank} \,\Gamma \le 1\) is necessary.

For \(P(a\mid x)>0\), let \(v_{a\mid x}\) be Bob's vector of conditional means. Its components satisfy
\begin{equation*}
(v_{a\mid x})_{y}-m_{By}=\frac{a\Gamma_{xy}}{1+am_{Ax}}.
\end{equation*}
If \(\Gamma=0\), one product label suffices. If \(\operatorname{rank} \Gamma=1\), all conditional vectors lie
on a line through \(m_{B}\). Let \(v_{1},v_{2}\) be the endpoints of their convex hull and write \(v_{a\mid
x}=\sum_{j} t_{j\mid a,x} v_{j}\). By no-signaling, \(\sum_{a} P(a\mid x)v_{a\mid x}=m_{B}\). Uniqueness of
the barycentric coordinates makes \(p_{j}=\sum_{a} P(a\mid x)t_{j\mid a,x}\) independent of \(x\). Draw label \(j\) with probability \(p_{j}\). Given \(j\), Alice outputs \(a\) with probability
\(\frac{P(a\mid x)t_{j\mid a,x}}{p_{j}}\) for supported outcomes and zero otherwise. Bob uses \(v_{j}\). This
reproduces Alice's marginal and every binary conditional distribution of Bob.

\subsection{Physical state family and its zero-cost members}
\label{sec:F.2}

In the computational basis, the state of Eq.~\eqref{eq:33} is
\begin{equation}
\label{eq:F.2}
\rho_{\zeta,\mu}=\frac{1}{4}\begin{pmatrix}
1-\zeta & 0 & 0 & 0 \\
0 & 1+\zeta+2\mu &-2\zeta & 0 \\
0 &-2\zeta & 1+\zeta-2\mu & 0 \\
0 & 0 & 0 & 1-\zeta
\end{pmatrix}.
\end{equation}
The middle block of \(4\rho_{\zeta,\mu}\) has trace \(2(1+\zeta)>0\) and determinant
\((1-\zeta)(1+3\zeta)-4\mu^{2}\). On \(0<\zeta \le \frac{1}{3}\), positivity is therefore equivalent to
Eq.~\eqref{eq:34}. Partial transposition of \(4\rho_{\zeta,\mu}\) moves the off-diagonal entries to the corner block, whose smaller
eigenvalue is \(1-3\zeta\). The other two diagonal entries are nonnegative by Eq.~\eqref{eq:34}, so the
two-qubit partial-transpose criterion \cite{Peres1996,Horodecki1996} gives separability.

For \(\zeta>0\), applying the rank criterion of Appendix~\ref{sec:F.1} to Eq.~\eqref{eq:36} gives
the zero-cost condition in Eq.~\eqref{eq:37}. An explicit realization on this zero-cost curve uses two equally
likely labels \(\varepsilon=\pm 1,\) with classical responses
\(u_{\varepsilon}=(\varepsilon,\mu)^{\mathsf{T}}\) and \(v_{\varepsilon}=(-\varepsilon
\zeta,-\mu)^{\mathsf{T}}\). These responses are admissible and reproduce the target local means. Since \(\mu^2=\zeta\), their correlation matrix is \(\overline{uv^{\mathsf{T}}}=-\zeta I_{2}\). This pure-CC protocol realizes the full target behavior. Together with the Werner result at \(\mu=0\) in
Appendix~\ref{sec:B.2}, this construction establishes the cost separation in Eq.~\eqref{eq:38}.

For fixed \(0<\zeta<\frac{1}{3}\), the physical boundary satisfies
\begin{equation}
\label{eq:F.3}
t_{\max}^{2}-\zeta=\frac{(1+\zeta)(1-3\zeta)}{4}.
\end{equation}
Thus \(0<\sqrt{\zeta}<t_{\max}\), and Eq.~\eqref{eq:37} gives zero cost only at \(t=\sqrt{\zeta}\).
At \(t=0\) and \(t=t_{\max}\), \(\operatorname{rank}\Gamma=2\), excluding a
zero-cost realization. Since the minimum cost is attained (Appendix~\ref{sec:A.1}), it is strictly positive at
both points. This proves Eq.~\eqref{eq:39a}. At \(\zeta=\frac{1}{3},\) the equality \(t_{\max}=\sqrt{\zeta}\)
places the zero-cost point on the physical boundary.

\subsection{Exact local-bias curve on the plotted interval}
\label{sec:F.3}

Fix \(\zeta=\frac{1}{9}\) and \(t=|\mu|\in [0,\frac{1}{2}]\). Simultaneous local output relabelings allow
\(\mu=t\). The case \(\delta=1\) already satisfies the bounds. For an arbitrary two-label protocol with
\(\delta<1\), use the effective responses \(\xi_{j},\chi_{j}\) defined in Eq.~\eqref{eq:B.2}. The matching
conditions give
\begin{equation}
\label{eq:F.4}
C=\mathbb{E}[(1,{\xi_{j}}^{\mathsf{T}})^{\mathsf{T}} (1,{\chi_{j}}^{\mathsf{T}})]=\tau M+\delta R,\quad R=\mathbb{E}[(1,{r_{j}}^{\mathsf{T}})^{\mathsf{T}} (1,{s_{j}}^{\mathsf{T}})],
\end{equation}
where
\begin{equation*}
M=\begin{pmatrix}
1 & 0 &-t \\
0 &-\zeta & 0 \\
t & 0 &-\zeta
\end{pmatrix}.
\end{equation*}
With at most two labels, \(\operatorname{rank} \,(C)\le 2\). A left null vector can be normalized as
\((c,n^{\mathsf{T}}),\) with \(\Vert n\Vert_{2}=1\). Its first-column equation implies \(|c|<1\).

For \(0\le t\le \frac{1}{3}\), reverse both of Bob's response coordinates. The target moment matrix becomes
\begin{equation*}
M_{+}=\begin{pmatrix}
1 & 0 & t \\
0 & \zeta & 0 \\
t & 0 & \zeta
\end{pmatrix}.
\end{equation*}
Contracting the transformed version of Eq.~\eqref{eq:F.4} with a left null vector \((c,n^{\mathsf{T}})\) and
its transpose gives
\begin{equation}
\label{eq:F.5}
0=\tau (c^{2}+2ctn_{Z}+\zeta)+\delta \mathbb{E}[(c+n\cdot r_{j})(c-n\cdot s_{j})].
\end{equation}
Both factors in the expectation lie in \([c-1,c+1]\). Since \(|c|<1\), their product is at least \(c^{2}-1\).
Completing the square and using \(|n_Z|\le1\) yields
\begin{equation}
\label{eq:F.6}
\delta-(1-\delta)\zeta+(1-\delta)^{2}t^{2}\ge0.
\end{equation}
For \(t^{2} \le \zeta\), the
left-hand side is strictly increasing in \(\delta \in [0,1].\) Its unique zero in this interval gives
\begin{equation*}
\delta \ge 1-\frac{2}{1+\zeta+\sqrt{(1+\zeta)^{2}-4t^{2}}},
\end{equation*}
which is the lower-bias part of Eq.~\eqref{eq:39b}.

For \(\frac{1}{3}<t\le \frac{1}{2}\), set \(D=t^{2}-\zeta>0\), \(h=\frac{\delta}{\tau}\) and
\(u(v)=(1,v_{X},v_{Z})^{\mathsf{T}}\). Here \(M\) is invertible and \(C\) is singular. Choose \(z\ne 0\) in
the right kernel of \(C\). By Eq.~\eqref{eq:F.4}, \(z=-hM^{-1} \mathbb{E}[u(r_{j})u(s_{j})^{\mathsf{T}} z]\).
Thus \(h>0\) and \(\max_{j}|u(s_{j})^{\mathsf{T}} z|>0\) on the support. Choose \(i_\ast\) to maximize \(\lvert u(s_i)^{\mathsf T}z\rvert\) over the support. Left-multiply the identity for \(z\) by \(u(s_{i_\ast})^{\mathsf T}\), take absolute values, and divide by \(\lvert u(s_{i_\ast})^{\mathsf T}z\rvert\) to obtain \(\frac{1}{h}\le \max_{i,j}|u(s_{i})^{\mathsf{T}} M^{-1}
u(r_{j})|\).

Put \(\alpha=\frac{D}{\zeta}\le 1+t\) and \(A(v)=v_{Z}^{2}+\alpha v_{X}^{2}+2t|v_{Z}|\). For a unit-disk
vector, \(v_{X}^{2}+2|v_{Z}|\le 2\) and \(\Vert v\Vert^{2} \le 1\) imply \(A(v)\le 1+2t\). Using the inverse
of \(M\) and \(2|ab|\le a^{2}+b^{2}\) gives
\begin{equation}
\label{eq:F.7}
\begin{gathered}
D|u(s)^{\mathsf{T}} M^{-1} u(r)|=|-\zeta+t(r_{Z}-s_{Z})+s_{Z} r_{Z}-\alpha s_{X} r_{X}| \\
\le \zeta+\frac{A(r)+A(s)}{2}\le 1+2t+\zeta.
\end{gathered}
\end{equation}
Hence \(h\ge \frac{D}{1+2t+\zeta}\), equivalently \(\delta \ge \frac{t^{2}-\zeta}{(1+t)^{2}}\), as claimed in
Eq.~\eqref{eq:39b}.

Both bounds are attained with two equally likely shared labels \(\varepsilon=\pm 1\). For each bias interval,
set \(\delta\) to its lower bound and choose \(d=\sqrt{\delta},\eta_{A}=\eta_{B}=1-d.\) For \(0\le t\le
\frac{1}{3}\), choose
\begin{equation*}
r_{\varepsilon}=s_{\varepsilon}=(0,\varepsilon)^{\mathsf{T}},\quad u_{\varepsilon}=\left( \varepsilon \frac{\sqrt{\tau \zeta}}{1-d},\frac{\tau t-d\varepsilon}{1-d}\right)^{\mathsf{T}},\quad v_{\varepsilon}=\left(-\varepsilon \frac{\sqrt{\tau \zeta}}{1-d},\frac{-\tau t-d\varepsilon}{1-d}\right)^{\mathsf{T}}.
\end{equation*}
The effective responses are \(\xi_{\varepsilon}=\left( \varepsilon \sqrt{\tau \zeta},\tau
t\right)^{\mathsf{T}}\) and \(\chi_{\varepsilon}=\left(-\varepsilon \sqrt{\tau \zeta},-\tau
t\right)^{\mathsf{T}}\). Eq.~\eqref{eq:F.4} reduces to \(\delta=\tau \zeta-\tau^{2} t^{2}\), which holds at
the first lower bound.

For \(\frac{1}{3}<t\le \frac{1}{2}\), choose
\begin{equation}
\label{eq:F.8}
\begin{gathered}
r_{\varepsilon}=(0,-1)^{\mathsf{T}},s_{\varepsilon}=(0,1)^{\mathsf{T}}, \\
u_{\varepsilon}=\left( \varepsilon \frac{\sqrt{\tau \zeta}}{1-d},t+\sqrt{t^{2}-\zeta}\right)^{\mathsf{T}},\quad v_{\varepsilon}=\left(-\varepsilon \frac{\sqrt{\tau \zeta}}{1-d},-t-\sqrt{t^{2}-\zeta}\right)^{\mathsf{T}}.
\end{gathered}
\end{equation}
The effective \(Z\) responses are \(\tau t-\delta\) for Alice and \(-(\tau t-\delta)\) for Bob.
Eq.~\eqref{eq:F.4} reduces to \((\tau t-\delta)^{2}=\tau \zeta+\delta\), which holds at the second lower
bound. In both protocols, the \(X\) components reproduce the required \(XX\) correlations. Equal label
probabilities ensure zero local \(X\) means and zero \(XZ\) and \(ZX\) correlations.

Throughout the two intervals, \(d<\frac{1}{3}\) and the squared classical \(X\) components satisfy
\(\frac{\tau \zeta}{(1-d)^{2}}<1\). On the first bias interval, the magnitudes of the classical
\(Z\) components are bounded by \(\frac{\tau t+d}{1-d}\le 1.\) On the second bias interval, their common
magnitude satisfies \(\left( t+\sqrt{t^{2}-\zeta}\right)<1\). The quantum response vectors have unit norm.
Thus both protocols satisfy the response constraints and all eight matching conditions at the corresponding
lower bounds. This establishes Eq.~\eqref{eq:39b} on \(0\le t\le \frac{1}{2}\).

The left and right derivatives at \(t=\frac{1}{3}\) are \(-\frac{3}{4}\) and \(\frac{3}{8}\), respectively,
giving the cusp in Fig.~\ref{fig:4}(b). The physical family extends beyond the proved interval, up to
\(t_{\max} \left( \frac{1}{9}\right)=\frac{2\sqrt{6}}{9}\).

\section{Source sampling and resource bounds}
\label{sec:G}

\vspace{-5pt}
\subsection{Classical baseline}
\label{sec:G.1}

At \(p=\frac{1}{3}\), the sampling target \(P=P_{\frac{1}{3}}^{\mathrm{samp}}\) in Eq.~\eqref{eq:40} has
\(E=+\frac{I_2}{3}\) after Bob's output flip of the Werner behavior with \(E=-\frac{I_2}{3}\).
This rank-two correlation matrix excludes pure CC with at most two labels by Eq.~\eqref{eq:B.1}.
Restricting the baseline to CC and QQ gives \(P=(1-\delta)P_{\mathrm{CC}}+\delta P\). Any \(\delta<1\) would imply
\(P=P_{\mathrm{CC}}\), so even coordinated choices require \(\delta=1\).

\subsection{Optimal resource use}
\label{sec:G.2}

Appendix~\ref{sec:B.2} at \(\zeta=\frac{1}{3}\) gives two equiprobable labels with
\(\eta_{A}=\eta_{B}=\frac{1}{2}\). The relabeling in Sec.~\ref{sec:7} transfers this protocol to the sampling
task. By AM--GM applied to the two activation probabilities and the bound \(\delta\ge\frac14\) from Theorem~\ref{thm:2},
\begin{equation}
\label{eq:G.1}
N_{\mathrm Q}\ge 1.
\end{equation}
This protocol attains equality in Eq.~\eqref{eq:G.1}, minimizing both resources at
\((\delta,N_{\mathrm{Q}})=(\frac{1}{4},1)\).

\subsection{Exact two-label approximation threshold}
\label{sec:G.3}

The sampling target is unbiased with correlation matrix \(\zeta I_{2}\), where \(\zeta=1-2p\in [0,1]\). For a
two-label classical behavior \(Q\), Appendix~\ref{sec:F.1} gives \(\operatorname{rank} \Gamma_{Q} \le 1\).
Choose \(v\in \ker \Gamma_{Q}\) with \(\Vert v\Vert_{1}=1\) and an input \(x\) and overall sign so that
\(v_{x} \ge \frac{1}{2}\). Bob privately samples \(y\) with probability \(|v_{y}|\) and returns
\(\operatorname{sgn} (v_{y})b\), omitting zero-weight inputs. For this fixed \(x\), the resulting binary pair
has covariance \((\Gamma_{Q} v)_{x}=0\) under \(Q\), so its law \(q\) is a product. Under the target its law
is unbiased with correlation \(c=\zeta v_{x} \ge \frac{\zeta}{2}\). This randomization and output relabeling
cannot increase the worst-case total variation distance.

Put \(R=\frac{\sqrt{1+c}}{2}\), so the two diagonal target probabilities are \(R^{2}\). For a product law,
\(\sqrt{q_{++}}+\sqrt{q_{--}}\le 1\) by Cauchy--Schwarz. Truncate each square root at \(R\). Since
\(\frac{1}{2}\le R\le 1\), the two resulting squares sum to at most \(R^{2}+(1-R)^{2}\), by convexity at fixed
sum. The deficits on the two diagonal outcomes sum to at least \(2R-1\). Using \(c\ge\frac{\zeta}{2}\) gives
\begin{equation}
\label{eq:G.2}
d(Q,P_{p}^{\mathrm{samp}})\ge\sqrt{1+\frac{\zeta}{2}}-1.
\end{equation}
For attainment, set \(e=\sqrt{1+\frac{\zeta}{2}}-1\) and \(m=(e,-e)^{\mathsf{T}}\). Draw a uniform shared sign \(\lambda=\pm1\). Conditional on \(\lambda\), the parties generate their
outputs independently with mean vectors
\begin{equation}
\label{eq:G.3}
u_{\lambda}=v_{\lambda}=m+\lambda \sqrt{\frac{\zeta}{2}}\,(1,1)^{\mathsf{T}}.
\end{equation}
The responses are admissible since \(e+\sqrt{\frac{\zeta}{2}}<1\). Both local mean vectors equal \(m\), and the correlators satisfy \((E_Q)_{xy}=m_xm_y+\frac{\zeta}{2}\). Substituting \(\zeta=4e+2e^{2}\) gives a correlator error of \(-2e\) on the diagonal and \(2e\) off the diagonal. After local sign flips making the two means
positive or zero, the four probability differences in the order \(++,+-,-+,--\) are
\((0,\frac{e}{2},\frac{e}{2},-e)\). Each input pair thus has distance \(e\), attaining the minimum in
Eq.~\eqref{eq:42}.

\phantomsection
\bibliography{references}

\end{document}